\documentclass[pre,aps,floats,superscriptaddress,floatfix]{revtex4}
\usepackage{amssymb,amsmath}
\usepackage{amsmath,amssymb}
\usepackage{graphicx}
\usepackage{psfrag}
\usepackage[normalem]{ulem}
\usepackage{xcolor}

\def\beq{\begin{equation}}
\def\eeq{\end{equation}}
\def\bea{\begin{eqnarray}}
\def\eea{\end{eqnarray}}

\newcommand{\blue}{\textcolor{black}}

\begin{document}
\title{Current fluctuations in non-interacting run-and-tumble particles in one-dimension} 
 
\author{Tirthankar Banerjee}
\affiliation{LPTMS, CNRS, Univ. Paris-Sud, Universit\'e Paris-Saclay, 91405 Orsay, France}
\affiliation{Instituut voor Theoretische Fysica, KU Leuven, 3001 Heverlee, Belgium}
\author{Satya N. \surname{Majumdar}}
\affiliation{LPTMS, CNRS, Univ. Paris-Sud, Universit\'e Paris-Saclay, 91405 Orsay, France}
\author{Alberto Rosso}
\affiliation{LPTMS, CNRS, Univ. Paris-Sud, Universit\'e Paris-Saclay, 91405 Orsay, France}
\author{Gr\'egory \surname{Schehr}}
\affiliation{LPTMS, CNRS, Univ. Paris-Sud, Universit\'e Paris-Saclay, 91405 Orsay, France}


\date{\today}
\begin{abstract}
We present a general framework to study the distribution of the flux through the origin up to time $t$, in a non-interacting one-dimensional system of particles with a step initial condition with a fixed density $\rho$ of particles to the left of the origin. We focus principally on two cases: (i) when the particles undergo diffusive dynamics (passive case) and (ii) run-and-tumble dynamics for each particle (active case). In analogy with disordered systems, we consider the flux distribution both for the annealed and the quenched initial conditions, for the passive and active particles. In the annealed case, we show that, for arbitrary particle dynamics, the flux distribution is a Poissonian with a mean $\mu(t)$ that we compute exactly in terms of the Green's function of the single particle dynamics. For the quenched case, we show that, for the run-and-tumble dynamics, the quenched flux distribution takes an anomalous large deviation form at large times $P_{\rm qu}(Q,t) \sim \exp\left[-\rho\, v_0\, \gamma \, t^2 \psi_{\rm RTP}\left(\frac{Q}{\rho v_0\,t} \right) \right]$, where $\gamma$ is the rate of tumbling and $v_0$ is the ballistic speed between two successive tumblings. In this paper, we compute the rate function $\psi_{\rm RTP}(q)$ and show that it is nontrivial. Our method also gives access to the probability of the rare event that, at time $t$, there is no particle to the right of the origin. For diffusive and run-and-tumble dynamics, we find that this probability decays with time as a stretched exponential, $\sim \exp(-c\, \sqrt{t})$ where the constant $c$ can be computed exactly.  
We verify our results for these large deviations by using an importance sampling Monte-Carlo method. 
\end{abstract}
\maketitle

\section{Introduction}

Current fluctuations in non-equilibrium open systems has been a major area of research in statistical physics over the last
few years \cite{bodineau-derrida-prl2004, derrida-gers, derrida-gers-sep, bertini-jstatphys2006,
bertini-landim-prl2005, prahofer-spohn-2002, prolhac-mallick-08,derrida-lecomte-wijland-pre2008,KM12}. The probability distribution of
the current of particles across a given point in space typically admits a large deviation form and the corresponding large deviation
function (rate function) is often interpreted as an analogue of a free energy in non-equilibrium systems. For example, it was shown in \cite{bodineau-derrida-prl2004} that for a large one dimensional system in contact with reservoirs at unequal densities at the two ends, there exists an additivity principle obeyed by the large deviation function, much like the free energy in equilibrium systems. This additivity principle has been exploited further to compute
cumulants of the current distribution \cite{prahofer-spohn-2002,prolhac-mallick-08,derrida-gers,derrida-gers-sep,bodineau-derrida-prl2004,bertini-jstatphys2006,bertini-landim-prl2005,derrida-lecomte-wijland-pre2008}. Several theoretical tools have been developed to study current fluctuations, 
notable among them is the Macroscopic Fluctuation Theory~\cite{bertini-jstatphys2006, derrida-gers, bodineau-derrida-prl2004,KM12} and the Bethe Ansatz~\cite{derrida-gers-sep}. Most of these studies have focused on driven diffusive systems, both interacting (as in the simple symmetric exclusion process) and non-interacting, typically in a one-dimensional setting.

Another inherently out-of-equilibrium system, much studied recently, is the so-called {\it active} run-and-tumble particle (RTP)~\cite{Berg_book,TC_2008,Martens2012,patterson-gopinath}, a new incarnation of the persistent random walk \cite{Sta87,Weiss}.
Such motion has been observed in certain bacteria such as E. Coli where the bacterium moves in straight runs, undergoes tumbling at the end of a run and chooses randomly a new direction for the next run~\cite{Berg_book,TC_2008,Martens2012,patterson-gopinath,Sta87,Weiss}. These motions are inherently out-of-equilibrium 
since they consume energy directly from the environment and self-propel themselves without any external force. There has been an enormous
amount of work concerning the collective properties of an assembly of such RTPs \cite{Fodor17,bechinger_active_2016,cates_motility-induced_2015,Cates_Nature,SEB_16}. Even at the single-particle level, RTP displays interesting behaviour and several single-particle observables have been studied recently. These include the position distribution for a free RTP \cite{Sta87,Martens2012,gradenigo}, non-Boltzmann 
stationary states for an RTP in a confining potential \cite{Cates_Nature, ad-sm-gh,Mallmin_18,Sevilla_19,HP95}, effects of disordered potentials \cite{Kardar-disorder}, first-passage properties \cite{km-ad-sm,maes,pierre-satya-greg,ADP_2014, A2015, MLW_86}, the distribution of the time at which an RTP reaches its maximum displacement \cite{anupam_rtp} and RTP subjected to stochastic resetting \cite{me-sm-reset, review_resetting}.

However, as far as we are aware, current fluctuations, even in a system of noninteracting RTPs have not been systematically studied. The purpose of this paper is to study the current fluctuations in the simplest setup 
where RTPs are noninteracting and initially confined on one-half of the real line (i.e. step-function initial condition). Such a setup was used before
by Derrida and Gerschenfeld for noninteracting diffusive particles \cite{derrida-gers} and they were able to compute the large-deviation form of the current
or flux $Q_t$ of particles through the origin {\it up to} time $t$. In this paper, we use exactly the same setup, but for a more general class of noninteracting particles, which includes both diffusive as well as run-and-tumble particles, and compute analytically the flux distribution up to time $t$. 

Thus the main observable of our interest is the flux $Q_t$ defined as the number of particles that crossed the origin (either from left or right) {\it up to} time $t$, starting from the step initial condition where the particles are uniformly distributed over only the left side of the origin. Let us denote its probability distribution by $P(Q,t) = {\rm Prob.}(Q_t=Q)$. Clearly $Q_t$ is a history dependent quantity, since it involves counting of all the crossings of the origin up to time $t$. Our exact results rely on a simple but crucial observation, valid for this special step initial condition: each particle, starting from the left side of the origin, that crosses the origin an even number of times up to time $t$ does not contribute to the flux $Q_t$. But if it crosses the origin an odd number of times, it contributes unity to the flux $Q_t$. Hence, the flux $Q_t$ is exactly equal to the number $N^+_t$ of particles present on the right side of 
the origin {\it at} time $t$, i.e. $Q_t = N^+_t$. Thus the history dependent observable $Q_t$ gets related, via this observation, to $N^+_t$ which is an instantaneous observable at time $t$. As we will see later, it is much easier to compute the probability distribution $P(N^+,t)={\rm Prob.}(N^+_t=N^+)$, rather than the distribution of $Q_t$ directly. Hence, knowing the distribution of $N^+_t$, we can compute the flux distribution from $P(Q,t) ={\rm Prob.}(N^+_t=Q)$. Note that this equivalence holds for arbitrary dynamics of the particles, for example it holds both for diffusive as well as RTP dynamics of the particles.

Based on this connection $Q_t = N^+_t$, we can apply our results for the flux distribution to another interesting problem. Consider for instance an ideal gas of noninteracting particles in 
a box. Imagine that the box is divided into two halves by a removable wall. Initially, all the particles are on the left half of the box and at $t=0$ we 
lift the wall and let the particles explore the full box freely. At time $t$, we take a snapshot of the system and observe the locations of 
the particles. Of course, on an average, one expects that, at long times, the particles will be uniformly distributed throughout the box. We can 
ask: what is the probability that, at time $t$, all the particles are again back to the left half of the box? Clearly this is an extremely rare event but
what is the probability of this event? How does it decay with time? But note that this is exactly the probability ${\rm Prob.}(N^+_t=0)=P(Q_t=0,t)$. Hence, our computation of the flux distribution with step initial condition gives access to the probability of this rare event (corresponding to all the particles coming back to the left half of the box at time $t$), in a one-dimensional setting.

The rest of the paper is organised as follows. In section~\ref{sec:model} we discuss the model and present a summary of our main results. Then in Sec.~\ref{general-setting-sec} we introduce the general setting and show how the single-particle Green's function plays the central role in the analysis. Next in Secs.~\ref{annealed-sec} and \ref{quen-sec}, we calculate the annealed and quenched averages, respectively, of the probability distribution of the flux for both diffusive and run-and-tumble particles.
Then in Sec.~\ref{numerics} we give the numerical verifications of our results and
finally in Sec.~\ref{conclu} we summarize and conclude. 
 
\section{The model and the main results}\label{sec:model}

\begin{figure}[htb]
\begin{center}
\includegraphics[width=8cm]{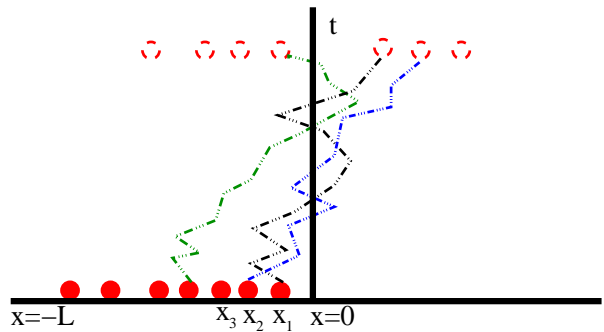}
\caption{Schematic representation of an initial realization with all particles  on the left of an arbitrary origin ($x=0$) on an infinite line $L \rightarrow \infty$. After time $t$ each particle undergoes some displacement depending on
the dynamics. The quantity of interest in our case is the number of particles on the right of the origin at time $t$.
}  \label{model}
\end{center}
\end{figure}

We consider a set of $N$ noninteracting particles initially distributed uniformly with a density $\rho$ on the negative real axis, as in Fig.~\ref{model}. Without loss of generality, we label the particles $i=1,2, \dots, N$ with $x_i(t)$ denoting the position of the $i$-th particle at time $t$. Each $x_i(t)$ evolves independently by a stochastic (or deterministic) evolution rule (the same law of evolution for each particle). For example, each particle can undergo independent Brownian motion. Alternatively, each particle can undergo independent RTP dynamics in one-dimension.  
This RTP dynamics for a single particle is defined as follows. 

\vspace{0.4cm}
\noindent {\bf RTP dynamics}. The position of a single RTP
$x(t)$ evolves via the Langevin equation
\begin{equation}
\frac{dx}{dt}= v_0\, \sigma(t)\, 
\label{RTP_evol.1}
\end{equation}
where $v_0$ is the intrinsic speed during a run and $\sigma(t)=\pm 1$ is a dichotomous 
telegraphic noise that flips from
one state to another with a constant rate $\gamma$. The effective noise $\xi(t)=v_0\, \sigma(t)$
is coloured which is simply seen by computing its autocorrelation function
\begin{equation}
\langle \xi(t) \xi(t')\rangle= v_0^2\, e^{-2\, \gamma\, |t-t'|}\, .
\label{autocorr.1}
\end{equation}
The time scale $\gamma^{-1}$ is the `persistence' time of a run that encodes the memory
of the noise. In the limit $\gamma\to \infty$, 
$v_0\to \infty$ but keeping the ratio $D_{\rm eff}= v_0^2/{2\gamma}$ fixed, the noise $\xi(t)$ reduces to
a white noise since
\begin{equation}
\langle \xi(t) \xi(t')\rangle= \frac{v_0^2}{\gamma}\, \left[\gamma\, e^{-2\gamma|t-t'|}\right]
\to 2D_{\rm eff}\, \delta(t-t')\, .
\label{autocorr.2}
\end{equation}
Thus in this so called `diffusive limit', the persistent random walker $x(t)$ reduces to an
ordinary Brownian motion. For finite $\gamma$ (i.e., persistence time scale of memory),
the RTP will be referred to as an ``active'' particle. In the diffusive limit, the active particle dynamics reduces
to an ordinary Brownian motion, which we refer to as a ``passive'' motion.

\begin{figure}[htb]
\begin{center}
\begin{minipage}{0.45\hsize}
\includegraphics[width=\hsize]{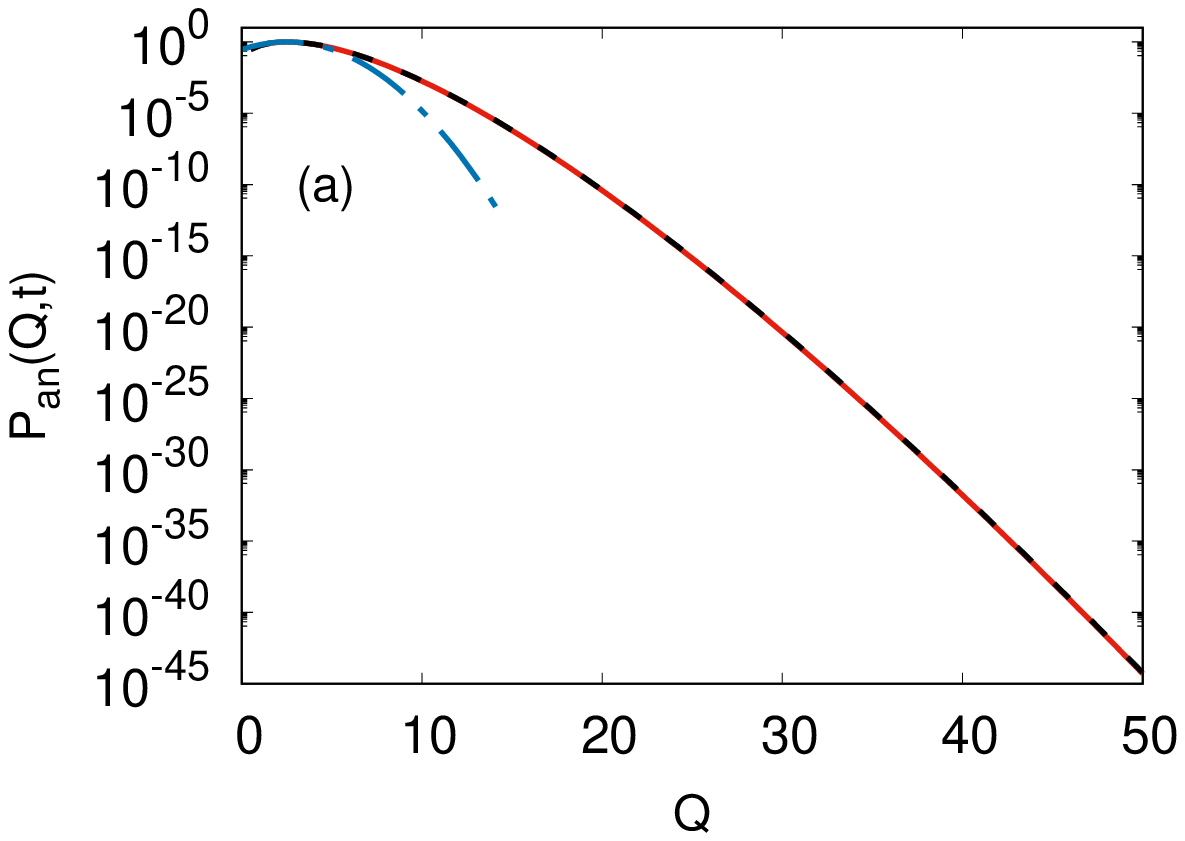}
\end{minipage}
\begin{minipage}{0.45\hsize}
\includegraphics[width=\hsize]{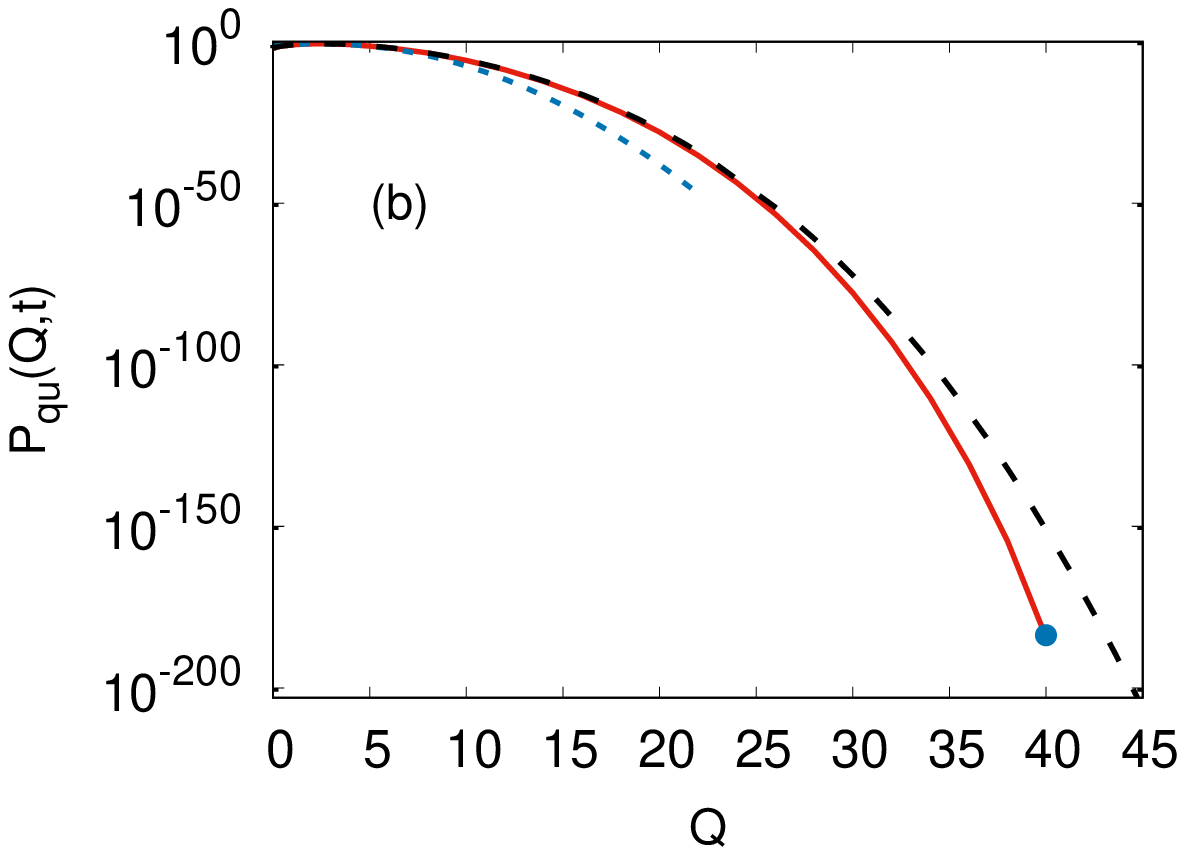}
\end{minipage}
\end{center}
\caption{{(a) Annealed case: Semi-log plot of $P_{\rm an}(Q,t)$ vs $Q$ for RTPs (red solid line) using Eqs.~(\ref{ac-mut}), (\ref{Poisson_largedev}) and (\ref{psi_an}) compared to the diffusive case (black dashed curve) given by Eq. (\ref{Poisson_largedev}) with $\mu(t)=\rho\sqrt{\frac{Dt}{\pi}}$ for $\rho = 1$ and $t=40$. For the RTP, we set $v_0=\gamma=1$, corresponding to an effective diffusion constant $D_{\rm eff}=v_0^2/(2\gamma)= 1/2$, while we use, accordingly, $D=1/2$ for diffusion. For such a (large) time where the scaling form (\ref{Poisson_largedev}) is expected to hold, the RTP and the diffusive cases are almost indistinguishable. Finally, the blue dashed curve corresponds to the typical Gaussian approximation with mean and variance $\mu(t)$ given in (\ref{ac-mut}). (b)~ Quenched case: Semi-log plot of $P_{\rm qu}(Q,t)$ vs $Q$ for the same set of parameters for RTPs (red solid line), obtained from Eqs.~(\ref{Uzt-exact}) and~(\ref{def_tildeI}), compared to the diffusive case (black dashed line) obtained from Eqs.~(\ref{phi-1eq}) and (\ref{fq-bas}). The blue dotted line corresponds to the Gaussian approximation with mean $\mu(t)$ given in (\ref{ac-mut}) and variance given in (\ref{var_qu}). While the active and passive cases remain
indistinguishable at small $Q$, the quenched distribution, at variance with the annealed one, carries a clear signature of activity  
at large $Q$.   For example, in the quenched case, the maximum possible flux for the RTP is $Q= \rho v_0 t = 40$ (large blue dot). 
}}\label{Pqu-compare}
\end{figure}

Given the stochastic dynamics of the individual particles, starting from the step initial condition, our main object of
interest is the flux $Q_t$ of particles through the origin up to time $t$. If a trajectory crosses the origin from left to right,
this will contribute a $+1$ to the net current while if it crosses from right to left, its contribution is $-1$. The flux $Q_t$ is thus
the net contribution to the current up to time $t$. Let us denote by $P(Q,t,\{x_i\})$ the probability distribution ${\rm Prob.}(Q_t=Q)$ 
for a given initial condition where $x_i$'s denote the initial positions of the particles at time $t=0$. 
Following Derrida and Gerschenfeld, 
the effect of the initial condition on the distribution can be studied in
two alternative ways, in analogy with the disordered systems where the realisation of a disorder plays an analogous
role as the initial condition in our problem. \blue{It was indeed argued in \cite{derrida-gers} that one has to distinguish between two different ways of 
averaging over the initial conditions: (i) the annealed average, where the probability distribution of the flux is averaged over all the realizations of the initial condition 
and (ii) the quenched average where the probability distribution is computed for the {\it typical} initial configurations.} Instead of considering the distribution $P(Q,t,\{x_i\})$ directly, it turns out to be convenient to consider its 
generating function $\langle e^{-p Q}\rangle_{\{x_i\}}$, where the angular brackets $\langle \cdots \rangle_{\{x_i\}}$ denote an average over the history, but with fixed initial condition $x_i$. The annealed and quenched averages are now defined as follows:
\begin{eqnarray}
&&\sum_{Q=0}^\infty e^{-p Q} \, P_{\rm an}(Q,t) \, =  \overline{\langle e^{-p Q}\rangle_{\{x_i\}}} \;, \label{def_ann} \\
&&\sum_{Q=0}^\infty e^{-p Q} \, P_{\rm qu}(Q,t) \,  = \exp{\left[ \overline{\ln \langle e^{-p Q}\rangle_{\{x_i\}}} \right]} \;, \label{def_quen}
\end{eqnarray}
where $\overline{\cdots}$ denotes an average over the initial conditions. Note that in this problem $Q_t$ is always an integer. As mentioned in the introduction, for the step initial condition, we can compute both $P_{\rm an}(Q,t)$ and $P_{\rm qu}(Q,t)$ \blue{(see Fig.~\ref{Pqu-compare} for a plot of these probability distributions)} for arbitrary dynamics of the particles by using the identity $Q_t = N^+_t$, where $N^+_t$ is the number of particles on the right side of the origin at time $t$. Indeed, the only quantity that enters the computation for independent particles is the single particle Green's function $G(x,x_0,t)$ denoting the probability density of finding the particle at position $x$ at time $t$, starting from $x_0$ at $t=0$. Let us first define a central object, that will appear in all our formulas 
\begin{eqnarray}\label{UzT}
U(z,t) = \int_0^\infty G(x,-z,t) \, dx \quad, \quad z\geq 0 \;,
\end{eqnarray}
obtained by integrating the Green's function over the final position, with the initial position fixed at $x_0=-z \leq 0$. If we can compute $U(z,t)$ for a given dynamics, we can express $P_{\rm an}(Q,t)$ and $P_{\rm qu}(Q,t)$ in terms of this central function $U(z,t)$. Our main results can now be summarised as follows.

\vspace*{0.4cm}
\noindent
{\bf Annealed case}: In this case, we show that $P_{\rm an}(Q,t)$ is always a Poisson distribution 
\begin{equation}\label{Pan-model-def}
P_{\rm an}(Q=n,t) = e^{-\mu(t)} \frac{\mu(t)^n}{n!} \;, \; n=0,1,2,\cdots \;,
\end{equation}
where 
\begin{equation}\label{muan-mod-def}
\mu(t) = \rho \, \int_0^{\infty} dz ~U(z,t) \;.
\end{equation} 
The mean and the variance of $Q_t$ are both given by $\mu(t)$, which 
can be explicitly evaluated for different types of particle motion. For example, for a Brownian motion with diffusion constant $D$, using $G(x,x_0,t) = e^{-(x-x_0)^2/(4 Dt)}/\sqrt{4 \pi D t}$ in Eq. (\ref{UzT}) we get 
\begin{eqnarray} \label{UzT_BM}
U(z,t) =  \frac{1}{2} {\rm erfc}\left(\frac{z}{\sqrt{4 D t}} \right) \quad, \quad {\rm and} \quad \mu(t) = \rho \sqrt{Dt/\pi} \;,
\end{eqnarray}
where ${\rm erfc}(z) = (2/\sqrt{\pi}) \int_z^\infty e^{-u^2} du$. Our result for $P_{\rm an}(Q=n,t)$ for the diffusive case is consistent with the result of Derrida and Gershenfeld \cite{derrida-gers} obtained by a different method.

In the case of the RTP dynamics, we find explicitly that at all $t$,
\begin{equation}\label{ac-mut}
\mu(t) = \frac{1}{2}\rho \, v_0 \, t\, e^{-\gamma t} \,[I_0(\gamma t) + I_1(\gamma t)],
\end{equation}
where $I_0(z)$ and $I_1(z)$ are modified Bessel functions of the first kind. Its asymptotic behaviours are given by
\begin{eqnarray} \label{asympt_mu}
\mu(t) \approx
\begin{cases}
&\dfrac{\rho \,v_0}{2}\, t \;, \;\;\; {\rm as}\;\;\; t \to 0 \;, \\
&\\
&\rho \sqrt{\dfrac{D_{\rm eff}\,t}{\pi}} \;, \;{\rm as}\; t \to \infty \;,
\end{cases}
\end{eqnarray}
where $D_{\rm eff} = v_0^2/(2 \gamma)$. Thus at late times, the RTP behaves like a diffusive particle with an effective diffusion
constant $D_{\rm eff}$. 

Note that the Poisson distribution in Eq. (\ref{Pan-model-def}) in the limit $Q \to \infty$, $\mu(t) \to \infty$, keeping
the ratio $Q/\mu(t)$ fixed, can be written in a large deviation form (using simply Stirling's formula)
\begin{eqnarray}\label{Poisson_largedev}
P_{\rm an}(Q,t) \sim \exp{\left[- \mu(t) \, \Psi_{\rm an} \left( \frac{Q}{\mu(t)}\right) \right]} \;,
\end{eqnarray}  
where the rate function $\Psi_{\rm an}(q)$ is universal, i.e., independent of the particle dynamics, and is given by
\begin{eqnarray} \label{psi_an}
\Psi_{\rm an}(q) = q \,\ln q - q + 1 \;, \; q \geq 0 \;.
\end{eqnarray}
It has the asymptotic behaviours
\begin{eqnarray} \label{asympt_psi_an}
\Psi_{\rm an}(q) \approx
\begin{cases}
& 1 \;, \;\; {\rm as} \;\; q \to 0 \\
& \dfrac{1}{2} (q-1)^2 \;, \;\; {\rm as} \;\; q \to 1 \\
& q \, \ln q \;, \;\; {\rm as} \;\; q \to \infty \;. 
\end{cases}
\end{eqnarray}
The quadratic behavior near the minimum at $q=1$ indicates typical Gaussian fluctuations for $Q$, with mean and variance both equal to $\mu(t)$. Note that the dependence on the particle dynamics in Eq. (\ref{Poisson_largedev}) enters only through the parameter $\mu(t)$ but the function $\Psi_{\rm an}(q)$ is universal. 

\begin{figure}[t]
\begin{center}
\begin{minipage}{0.4\hsize}
\includegraphics[width=\hsize]{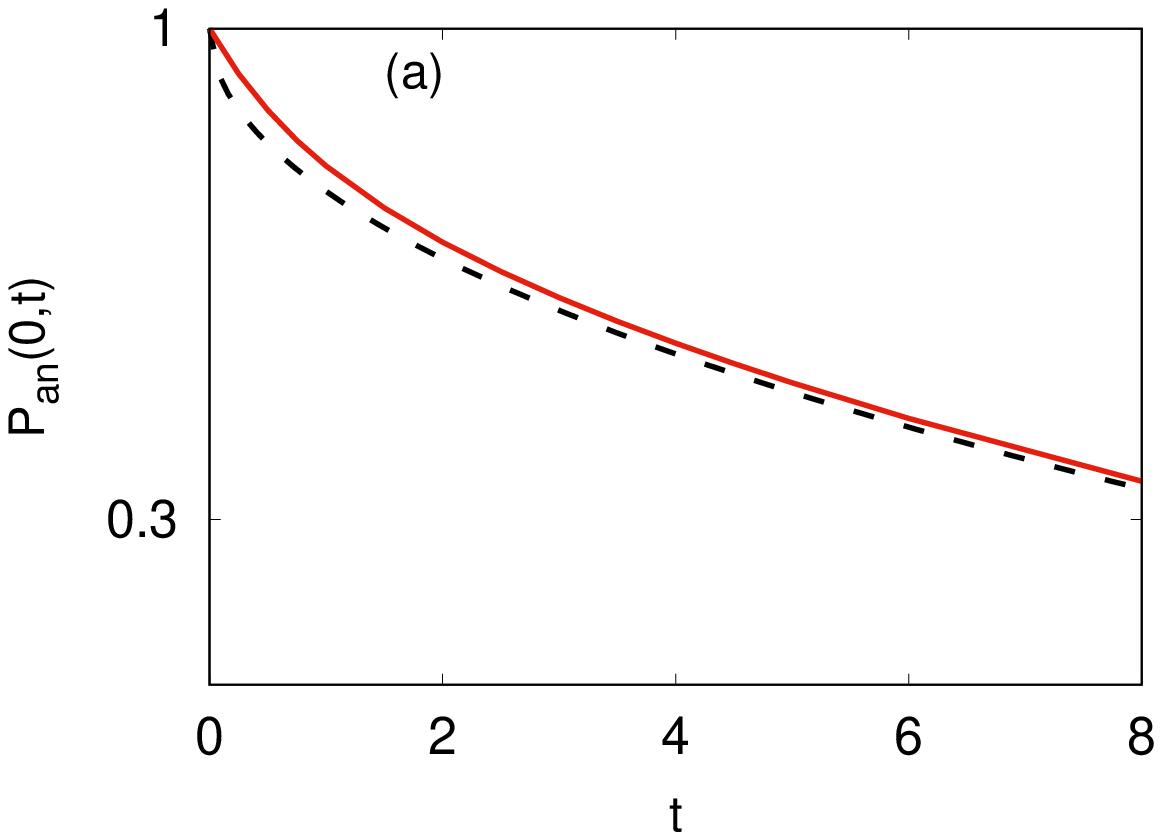}
\end{minipage}
\begin{minipage}{0.4\hsize}
\includegraphics[width=\hsize]{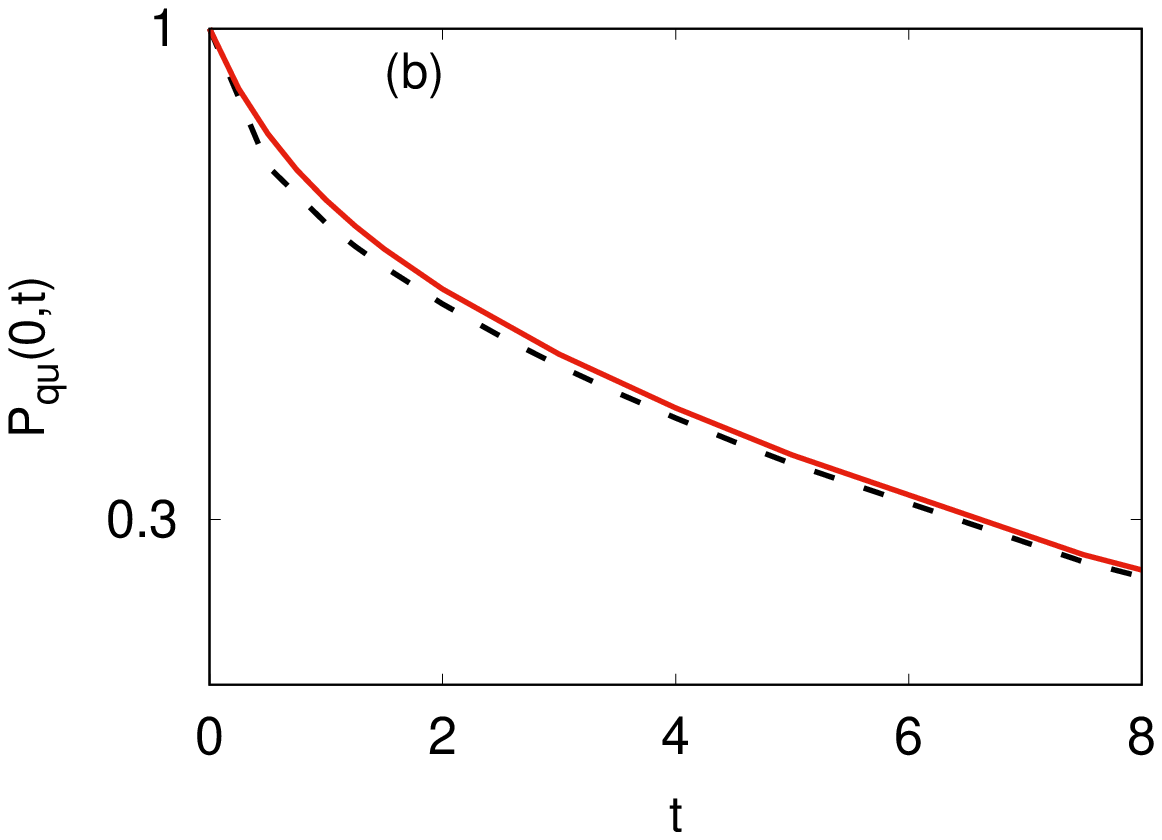}
\end{minipage}
\end{center}
\caption{{(a) Semi-log plot of $P_{\rm an}(Q=0,t) = e^{-\mu(t)}$ vs $t$ for an RTP (red solid line) with $\mu(t)$ given in Eq.~(\ref{Pq0})
 compared to the diffusive case (black dashed line) corresponding to $\mu(t)=\rho \sqrt{\frac{D t}{\pi}}$, for a density $\rho=1$ in both cases. For the RTP, we set $v_0=\gamma=1$, corresponding to an effective diffusion constant $D_{\rm eff}=v_0^2/(2\gamma)= 1/2$, while we set, accordingly, $D=1/2$ in the case of diffusion. (b) Semi-log plot of $P_{\rm qu}(Q=0,t)$ vs $t$ as given in Eq.~(\ref{P_neq0_qu}) both for RTPs (red solid line) using the result for $U(z,t)$ in (\ref{Uzt-exact}) and for Brownian particles (black dashed line) for which $U(z,t)$ is given in~(\ref{UzT_BM}). In both annealed and quenched cases, the zero-net flux probabilities for active and passive systems differ at short times but do coincide in the large time limit.}}\label{Pan-qu-compare}
\end{figure}

We also note that, from our general result in Eq. (\ref{Pan-model-def}), it follows that 
\begin{eqnarray}\label{Pq0}
{\rm Prob.}(N_t^+ =0) \Big|_{\rm an} = P_{\rm an}(Q=0,t) = e^{-\mu(t)} \;.
\end{eqnarray}  
This result is valid for all $t$ and gives the probability of the rare event that all the particles are back on the left side of the origin at time $t$, as discussed in the introduction. {In Fig.~\ref{Pan-qu-compare} (a) we show a plot of $P_{\rm an}(Q=0,t)$ as a function of time, both for RTP and for diffusive particles.}

\vspace*{0.4cm}
\noindent
{\bf Quenched case}: In this case, the generating function in Eq. (\ref{def_quen}), for arbitrary single particle dynamics, can again be expressed in terms 
of the central function $U(z,t)$ (\ref{UzT}) as follows
\begin{eqnarray}\label{P_qu_gen}
\sum_{Q=0}^\infty P_{\rm qu}(Q,t) e^{-p Q} \, = \exp{\left[\rho \int_0^\infty dz \, \ln{\left[1-(1-e^{-p})U(z,t)\right]} \right]} \;.
\end{eqnarray} 
The quenched cumulants of $Q$ can then be extracted and expressed in terms of $U(z,t)$. For instance, 
the quenched mean and the variance of $Q$ are given by  
\begin{eqnarray}
&&\langle Q \rangle_{\rm qu} = \rho \, \int_0^\infty U(z,t) \, dz \;,  \label{mean_qu} \\
&&\sigma_{\rm qu}^2 = \langle Q^2 \rangle_{\rm qu} - \langle Q \rangle^2_{\rm qu} = \rho \int_0^\infty U(z,t)(1-U(z,t))\, dz \;. \label{var_qu}
\end{eqnarray}
In addition, expanding the right hand side (rhs) in powers of $e^{-p}$ and matching with the left hand side (lhs), one can in principle obtain $P_{\rm qu}(Q,t)$ for any integer $Q$ as a functional of $U(z,t)$. For example, 
\begin{eqnarray}\label{P_neq0_qu}
{\rm Prob.}(N_t^+ = 0) \Big |_{\rm qu} = P_{\rm qu}(Q=0,t) = \exp{\left[\rho \int_0^\infty \, \ln{\left[1-U(z,t)\right]} \, dz\right]} \;,
\end{eqnarray}
which is valid for all times $t \geq 0$. However, the formula gets more complicated for higher values of $Q$. {In Fig.~\ref{Pan-qu-compare} (b) we show a plot of $P_{\rm qu}(Q=0,t)$ as a function of time, both for RTP and for diffusive particles.}

For the full quenched distribution, we first consider the diffusive motion of the particles. In this case, we obtain, in the scaling limit $Q \to \infty$, $t \to \infty$ keeping the ratio $Q/\sqrt{t}$ fixed, the same large deviation form as Derrida and Gerschenfeld~\cite{derrida-gers},
\begin{equation}\label{P-F-diff-largeQ}
P_{\rm qu}(Q,t) \sim \exp\left[-\rho \sqrt{D t} ~ \Psi_{\rm diff}\left(\frac{Q}{\rho \sqrt{D t}}\right)\right] \;,
\end{equation}
where the rate function $\Psi_{\rm diff}(q)$ has the following precise asymptotics
\begin{eqnarray} \label{psi_diff_asympt}
\Psi_{\rm diff}(q) \approx
\begin{cases}
&\overline{\alpha} - q + q \ln(q/\overline{\beta}) \;, \; {\rm as} \; q \to 0 \\
& \\
& \sqrt{\dfrac{\pi}{2}} \left(q-\dfrac{1}{\sqrt{\pi}} \right)^2 \;, \; {\rm as} \; q \to 1/\sqrt{\pi} \\
& \\
& \frac{1}{12} q^3 \;, \; {\rm as} \; q \to \infty \;,
\end{cases} 
\end{eqnarray}
with the two constants $\overline{\alpha}$ and $\overline{\beta}$ given explicitly by
\begin{eqnarray} 
&&\overline{\alpha} = -2 \int_0^\infty dz \, \ln\left(1 - \frac{1}{2} \, {\rm erfc}(z) \right) = 0.675336 \ldots \label{C} \\
&&{\overline{\beta}} = \int_0^\infty dz \, \frac{{\rm erfc}(z)}{1-\frac{1}{2} {\rm erfc}(z)} = 0.828581 \ldots \label{A} \;.
\end{eqnarray}
Note that the large $q$ behavior $\Psi_{\rm diff}(q)  \approx q^3/12$ coincides with the result of Derrida and Gerschenfeld \cite{derrida-gers} obtained
by a different method. The small $q$ behavior was not investigated in Ref. \cite{derrida-gers}. Taking the $q \to 0$ limit in Eq. (\ref{P-F-diff-largeQ}) and using the small $q$ behavior in the first line of Eq. (\ref{psi_diff_asympt}) implies that for large $t$ 
$P_{\rm qu}(Q=0,t) \sim \exp \left[ - \overline{\alpha} \, \rho \sqrt{D\,t}\right]$ where the constant $\overline{\alpha}$ is given in Eq. (\ref{C}). In fact, this result is valid not just at large time but at all times. Indeed, substituting $U(z,t) = (1/2) {\rm erfc}(z/\sqrt{4Dt})$ in our general formula (\ref{P_neq0_qu}), it follows that at all times $t \geq 0$,
\begin{eqnarray} \label{P_Q0}
P_{\rm qu}(Q=0,t)\Big \vert_{\rm diff} = \exp \left[ - \overline{\alpha} \, \rho \sqrt{D\,t}\right] \;.
\end{eqnarray}
Interestingly, exactly the same rate function $\Psi_{\rm diff}(q)$ also appeared in the completely different context, namely as a large deviation function characterising the distribution of the number of eigenvalues (in a disk of radius $R$) of a complex Ginibre ensemble of $N\times N$ Gaussian random matrices \cite{castillo,bertrand}.

We then consider the quenched flux distribution for the RTP dynamics. First, we show that, for small $Q \ll \sqrt{t}$, the flux distribution decays as a stretched exponential at late times, e. g. $P_{\rm qu}(Q=0,t) $ is given, for large $t$, by
\begin{eqnarray}\label{P_Q0-1_RTP}
P_{\rm qu}(Q=0,t) \Big \vert_{\rm RTP} \sim \exp \left[ - \overline{\alpha} \, \rho \sqrt{D_{\rm eff}\,t}\right] \;,
\end{eqnarray}
where $\bar{\alpha}$ is the same constant as in Eq. (\ref{C}) and $D_{\rm eff} = v_0^2/(2 \gamma)$. One of the main results of this analysis is to find a new scaling limit $Q \to \infty$, $t \to \infty$, keeping the ratio $Q/(\rho\,v_0 t)$ fixed where the quenched distribution admits a large deviation form [quite different from the diffusive case in Eq. (\ref{P-F-diff-largeQ})]
\begin{equation}\label{ac-sig}
P_{\rm qu}(Q,t) \sim \exp\left[-\rho \, v_0 \, \gamma \, t^2 \, \Psi_{\rm RTP}\left(\frac{Q}{\rho \, v_0 \,t}\right)\right],
\end{equation}
where the rate function $\Psi_{\rm RTP}(q)$ is given explicitly by
\begin{equation}\label{rtp-ldf-model}
\Psi_{\rm RTP}(q)=q-\frac{q}{2}\sqrt{1-q^2}-{\rm sin}^{-1}\left[ \sqrt{\frac{1-\sqrt{1-q^2}}{2}} \right]\;, \quad 0 \leq q \leq 1 \;.
\end{equation}
The rate function has the asymptotic behavior
\begin{eqnarray}\label{asympt_RTP}
\Psi_{\rm RTP}(q) \approx
\begin{cases}
& \dfrac{q^3}{6} \quad,\quad \quad \quad \; q\rightarrow 0 \,\\ 
& \\
& 1-\dfrac{\pi}{4} \quad,\quad ~ ~ ~ q = 1 \;.
\end{cases}
\end{eqnarray}
One consequence of our result is the prediction of the probability of the rare event that the flux $Q$ up to time
$t$ achieves its maximum possible value, namely $Q = \rho \, v_0\, t$ -- this corresponds to the case where all
the particles move ballistically to the right up to time $t$. We find that the probability of this rare event is given by
\begin{eqnarray} \label{P_qu_max}
P_{\rm qu}(Q = \rho \, v_0 \, t,t) \approx \exp\left[-\left( 1 - \frac{\pi}{4}\right) \rho \, v_0 \, \gamma \, t^2 \right] \;.
\end{eqnarray}
Such a faster than exponential decay for the probability of this rare event is a nontrivial prediction of our
theory. 

\section{The general setting and the single-particle Green's function}\label{general-setting-sec}

We start with a step initial condition where $N$ particles are initially located on the negative half line at positions $\{x_1, x_2, \cdots, x_N \}$ where 
all $x_i <0$. As stated before, for this step initial condition, the flux $Q_t$ up to time $t$ is identical in law to the number of particles $N^+_t$ to the
right of the origin at time $t$. Let us introduce an indicator function ${\cal I}_i(t)$ 
such that ${\cal I}_i(t)=1$ if the $i{\rm th}$ particle
is to the right of the origin at time $t$, else ${\cal I}_i(t)=0$. Hence we have
\begin{equation}\label{N+defn}
 N_t^+= \sum_{i=1}^N {\cal I}_i(t) \;.
\end{equation}
For fixed $x_i$'s the flux distribution is then given by
\begin{eqnarray}\label{basic-defn}
P(Q,t,\{ x_i\}) = {\rm Prob.}(N_t^+ = Q) = \left \langle \delta \left[Q-\sum_{i=1}^N {\cal I}_i(t)\right]\right \rangle_{\{x_i\}} \;,   
\end{eqnarray}
where the angular brackets $\langle \cdots \rangle_{\{x_i\}}$ denote an average over the history, but with fixed initial condition $x_i$. Taking the Laplace transform on both sides of Eq. (\ref{basic-defn}) gives
\begin{equation}\label{lpbasic}
 \sum_{Q=0}^{\infty} e^{-pQ} P(Q,t,\{x_i\}) = \langle e^{-pQ}\rangle_{\{x_i\}} = \left \langle \exp[{-p \sum_{i=1}^N {\cal I}_i(t)}]\right \rangle_{\{x_i\}} \;.
 \end{equation}
Since the ${\cal I}_i$ can only take the values $0$ or $1$, one has the identity $e^{-p {\cal I}_i} = 1 -(1-e^{-p}){\cal I}_i$. Inserting this identity in Eq. (\ref{lpbasic}) and using the independence of the random variables ${\cal I}_i$'s we get
 \begin{equation}\label{sq-Ii}
  \langle e^{-pQ}\rangle_{\{x_i\}} = \prod_{i=1}^N\left[1- (1-e^{-p})\langle {\cal I}_i(t) \rangle_{\{x_i\}} \right],
 \end{equation}
where the right hand side (r.h.s.) implicitly depends on the $x_i$'s. The average $\langle {\cal I}_i(t) \rangle_{\{x_i\}}$ is just the probability that the $i{\rm th}$ particle is to the right of the origin at time $t$, starting initially at $x_i$ and hence we have 
\begin{equation}\label{I-propagator}
  \langle {\cal I}_i(t) \rangle_{\{x_i\}} = \int_0^{\infty} G(x,x_i,t) dx = U(-x_i,t) \;, \quad x_i < 0 \;,
 \end{equation}
where $G(x,x_i,t)$ is the single-particle Green's function, i.e., the propagator for a particle to reach $x$ at time $t$, starting initially at $x_i<0$. Note that $U(z,t)$ is defined in Eq. (\ref{UzT}) and corresponds to the probability that a particle is on the positive side of the origin at time $t$, starting initially at $-z<0$. Inserting Eq.~(\ref{I-propagator}) into Eq.~(\ref{sq-Ii}), one obtains
\begin{equation}\label{propagator}
  \langle e^{-pQ}\rangle_{\{x_i\}}= \prod_{i=1}^N\left[1- (1-e^{-p})U(-x_i,t)\right]  \;, \quad x_i < 0 \;, \quad \forall i = 1, \cdots, N \;.
 \end{equation}
This Eq.~(\ref{propagator}) is general, i.e., valid for
{\em any} set of non-interacting particles undergoing a common dynamics in one-dimension. The information about the dynamics
is entirely encoded in the function $U(z,t)$. 

For instance, for simple diffusion, the single-particle Green's function is given by 
\begin{equation}\label{diff-pro2}
G(x,x_i,t) = \frac{1}{\sqrt{4\pi D t}}{{\rm exp}\left[-\frac{(x-x_i)^2}{4Dt}\right]} \;,
\end{equation}
which gives $U(z,t) = (1/2) {\rm erfc}(z/\sqrt{4Dt})$. For the RTP, on the other hand, the Green's function is known explicitly~\cite{Weiss,me-sm-reset} 
\begin{equation}\label{ac-fullpd3}
G(x,x_i,t) = \frac{e^{-\gamma t}}{2}\left\{ \delta(x-x_i-v_0t) + \delta(x-x_i+v_0t) +\frac{\gamma}{v_0}\left[I_0(\omega) + \frac{\gamma t I_1(\omega)}{\rho} \right] \Theta(v_0t-|x-x_i|) \right\},
\end{equation}
where $\omega$ is given by
\begin{eqnarray}\label{def_omega}
\omega = \frac{\gamma}{v_0}\sqrt{v_0^2t^2 -(x-x_i)^2} \;.
\end{eqnarray}
In Eq. (\ref{ac-fullpd3}), $\Theta(z)$ is the Heaviside Theta function, and $I_0(\omega)$ and $I_1(\omega)$ are modified Bessel functions. Computing $U(z,t) = \int_0^\infty G(x,-z,t)\,dx$ explicitly using Eq. (\ref{ac-fullpd3}) is complicated. It is however much more useful, as we will see later, to work with the Laplace transform of $G(x,x_i,t)$ with respect to $t$, which has a much simpler expression, namely 
\begin{equation}\label{ac-Lap3}
\tilde{G}(x,x_i,s) = \int_0^{\infty} dt~ e^{-st} G(x,x_i,t) = \frac{\lambda(s)}{2s}e^{\lambda(s)\,|x-x_i|} \;, \quad \lambda(s) = \frac{\sqrt{s(s+2\gamma)}}{v_0}  \;.
\end{equation}

The relation in Eq. (\ref{propagator}) is the central result of this section and we will analyse the annealed and the quenched cases separately in the next two sections.

\section{Flux distribution in the annealed case}\label{annealed-sec}

The annealed distribution $P_{\rm an}(Q,t)$ is defined in Eq. (\ref{def_ann}) where the $\overline{\cdots}$ denotes an average over
the initial conditions. Performing this average in Eq. (\ref{propagator}) gives
\begin{equation}\label{ann-G-avg}
\langle\overline{e^{-pQ}\rangle_{\{x_i\}}}= \prod_{i=1}^N\left[1- (1-e^{-p})\, \overline{U(-x_i,t)}\right] \;,
\end{equation}
where $U(-x_i,t)$ is defined in Eq. (\ref{I-propagator}). To perform the average over the initial conditions with a fixed uniform density $\rho$, 
we assume that each of the $N$ particles is distributed independently and uniformly over a box $[-L,0]$ and then eventually take the limit 
$N \to \infty$, $L \to \infty$ keeping the density $\rho = N/L$ fixed. For this uniform measure, each $x_i$ is uniformly distributed in the box
$[-L,0]$. Using the independence of the $x_i$'s we then get 
\begin{equation}\label{x_i-avg}
\langle\overline{e^{-pQ}\rangle_{\{x_i\}}} = \prod_{i=1}^N \left[1- (1-e^{-p}) \int_{-L}^{0} U(-x_i,t) \frac{dx_i}{L} \right] = \left[1- \frac{1}{L}(1-e^{-p}) \int_0^L U(z,t) dz \right]^N \;,
\end{equation}
where, in the last equality, we made the change of variable $z=-x_i$. Taking now the limit $N \to \infty$, $L \to \infty$ keeping $\rho = N/L$ fixed gives
\begin{eqnarray}\label{lap-pan}
\sum_{Q=0}^{\infty} e^{-pQ} P_{\rm an}(Q,t) = \langle\overline{e^{-pQ}\rangle_{\{x_i\}}} = \exp\left[-\mu(t) ~ (1-e^{-p})\right] \;, \quad {\rm where} \;\quad \mu(t)=  \rho \int_0^{\infty} dz ~ U(z,t) \;.
\end{eqnarray}
By expanding $\exp\left[-\mu(t) ~ (1-e^{-p})\right]$ in powers of $e^{-p}$ and comparing to the left hand side, we see that $Q$ can take only integer values $Q=n=0,1,2,\cdots$ and the probability distribution is simply a Poisson distribution with mean $\mu(t)$ as given in Eqs. (\ref{Pan-model-def}) and (\ref{muan-mod-def}). 

This Poisson distribution, in the annealed case, is thus universal, i.e., holds for any dynamics. The details of the dynamics is encoded in the single
parameter $\mu(t)$ which can be computed explicitly for different types of dynamics. For example, for diffusing particles, using the explicit expression for the Brownian propagator, we get $U(z,t) = (1/2) {\rm erfc}(z/\sqrt{4Dt})$ and, hence, $\mu(t) = \rho \sqrt{Dt/\pi}$ as mentioned in Eq. (\ref{UzT_BM}). In contrast, for the RTP dynamics, $\mu(t)$ is nontrivial. As discussed earlier, computing $U(z,t)$ from the Green's function in Eq. (\ref{ac-fullpd3}) is difficult. Consequently, calculating $\mu(t) = \rho\int_0^\infty U(z,t)\,dz$ is also hard. However, it turns out that its Laplace transform is much easier to manipulate, due to the simple nature of the formula in Eq. (\ref{ac-Lap3}). The Laplace transform of $\mu(t)$ is given by
\begin{equation}\label{mus-RTP}
\tilde{\mu}(s) = \int_{0}^{\infty} dt~ e^{-st}~ \mu(t) =  \rho \int_0^\infty  dz  ~\tilde{U}(z,s)  \;, \quad {\rm with} \quad \tilde U(z,s)  = \int_{0}^{\infty} dt~ e^{-st}~ U(z,t) \;,
\end{equation}
where we have used the relation $\mu(t) = \rho \int_0^\infty U(z,t)\,dz$. The Laplace transform of $U(z,t)$ can be computed as follows
\begin{eqnarray}\label{LaplaceU_1}
\tilde U(z,s)  = \int_{0}^{\infty} dt~ e^{-st}~ U(z,t) = \int_0^\infty dt \, e^{-st} \, \int_0^\infty G(x,-z,t) \, dx \;.
\end{eqnarray}
Exchanging the integrals over $x$ and $t$, and using the relation in Eq. (\ref{ac-Lap3}) and integrating over $x$ we get
\begin{eqnarray}\label{LaplaceU_2}
\tilde U(z,s) = \frac{e^{-\lambda(s) z}}{2s} \;, \quad {\rm where} \quad \lambda(s) = \frac{\sqrt{s(s+2\gamma)}}{v_0} \;.
\end{eqnarray}
Inserting this relation in Eq. (\ref{mus-RTP}) and performing the integral over $z$, we get
\begin{equation}\label{mu-ac-lap}
\tilde{\mu}(s) =  \frac{1}{2 s \, \lambda(s)} =  \frac{v_0}{2s\sqrt{s(s+2\gamma)}} \;.
\end{equation}
This Laplace transform can be explicitly inverted, yielding the result in Eq. (\ref{ac-mut}).

\section{Flux distribution in the quenched case}\label{quen-sec}

As stated in Section \ref{sec:model}, the quenched flux distribution is defined as
\begin{equation}\label{quenched-definition}
\sum_{Q=0}^{\infty} P_{\rm qu} (Q,t) e^{-pQ} = \exp \left[\overline{{\ln}\left[\langle e^{-pQ}\rangle_{\{x_i\}} \right]} \right] ,
\end{equation}
where $\overline{\cdots}$ once again represents an average over the initial positions $\{x_i\}$. Our starting point is again Eq. (\ref{propagator}). Taking the logarithm on both sides of (\ref{propagator}) gives
\begin{equation}\label{Nlog-quenched}
{\ln}\left[\langle e^{-pQ}\rangle_{\{x_i\}} \right] = \sum_{i=1}^N {\ln}\left[1-(1-e^{-p})U(-x_i,t) \right].
\end{equation}
We now perform the average over the initial positions, as in the annealed case, i.e., choosing each $x_i$ independently and uniformly from the box $[-L,0]$ and finally taking the limit $N \to \infty$, $L \to \infty$ keeping $\rho = N/L$ fixed. This gives
\begin{equation}\label{Nlog-qu-avg}
\overline{{\rm log}\left[\langle e^{-pQ}\rangle_{\{x_i\}} \right]}= \frac{N}{L}\int_{-L}^0 dx_i ~{\ln}\left[1-(1-e^{-p})U(-x_i,t) \right] \longrightarrow \rho \int_0^\infty dz\, \ln \left[ 1 - (1-e^{-p}) U(z,t)\right] \;.
\end{equation} 
Therefore the Laplace transform of the quenched flux distribution is given by
\begin{equation}\label{Pqu2}
\sum_{Q=0}^{\infty} P_{\rm qu} (Q,t) e^{-pQ}  = \exp\left[I(p,t)\right] \;,
\end{equation}
where 
\begin{eqnarray}\label{def_Ip}
I(p,t) = \rho \int_0^\infty dz\, \ln \left[ 1 - (1-e^{-p}) U(z,t)\right] \;.
\end{eqnarray}
Before extracting the full distribution $P_{\rm qu} (Q,t)$ from this Laplace transform, it is useful to study first the asymptotic behaviors of $I(p,t)$ 
in the two limits : (i) $p \rightarrow 0$ and (ii) $p \rightarrow \infty$. 
\begin{itemize}
\item[$\bullet$] $p\to 0$ limit: Expanding $e^{-p}$ in powers of $p$ in Eq. (\ref{def_Ip}), we get
\begin{equation}\label{Ip0}
I(p,t) = - p ~\rho \int_0^{\infty} dz~U(z,t) + \frac{p^2}{2} ~\rho \int_0^{\infty} dz~ U(z,t)\left[1-U(z,t) \right] + {\cal O}(p^3) \;.
\end{equation} 
Substituting this in Eq. (\ref{Pqu2}) and expanding both sides in powers of $p$ we immediately get the mean and the variance of the flux $Q_t$ for the quenched case
as stated in Eqs. (\ref{mean_qu}) and (\ref{var_qu}) respectively.    
 
\item[$\bullet$] $p\to \infty$ limit: In this case we expand $I(p,t)$ in Eq. (\ref{def_Ip}) in powers of $e^{-p}$. The two leading terms are given by
\begin{equation}\label{Ip-pinf1}
I(p,t)  = A(t) + B(t) e^{-p} + {\cal O}(e^{-2p}) \;,
\end{equation}
where
\begin{eqnarray}\label{atbt-pinf1}
A(t) &=& \rho \int_0^{\infty} {\ln} [1-U(z,t)] dz \\ 
B(t) &=&  \rho \int_0^{\infty} \frac{U(z,t)}{1-U(z,t)} dz \;.
\end{eqnarray}
Substituting this expansion (\ref{Ip-pinf1}) on the rhs of Eq. (\ref{Pqu2}) and matching the powers of $e^{-p}$ on both sides of Eq. (\ref{Pqu2}) immediately gives
\begin{eqnarray}
P_{\rm qu}(Q=0,t) &=& e^{A(t)} = \exp{\left[\rho\int_0^\infty \ln(1-U(z,t))dz\right]} \label{Pqu_eq_0} \\
P_{\rm qu}(Q=1,t) &=& B(t) \, e^{A(t)}  \;. \label{Pqu_eq_1}
\end{eqnarray}
The first line yields the general result mentioned in Eq. (\ref{P_neq0_qu}).

\end{itemize}

These results so far are quite general, i.e., they hold for any dynamics -- the dependence on the dynamics comes
only through the function $U(z,t)$. In the following, we focus on two interesting dynamics, namely the diffusive and 
the RTP and extract the large time behavior of $P_{\rm qu}(Q,t)$ using Eqs. (\ref{Pqu2}) and (\ref{def_Ip}).

\subsection{$P_{\rm qu}(Q,t)$ for simple diffusion}\label{quen-sec-diff}

In this case, using the explicit expression $U(z,t) = (1/2) {\rm erfc}(z/\sqrt{4Dt})$, we get from Eq. (\ref{def_Ip}) 
\begin{equation}\label{I-phi-correspondence}
I(p,t) = \rho \sqrt{ 4 D t} \int_0^{\infty} dz~ {\ln}\left[1-\frac{1}{2}(1-e^{-p}){{\rm erfc}(z)}\right]
= -\rho \sqrt{D t}~ \phi (p),
\end{equation}
where 
\begin{equation}\label{phi-1eq}
\phi (p) = -2 \int_0^{\infty} dz~ {\ln}\left[1-\frac{1}{2}(1-e^{-p}){{\rm erfc}(z)}\right ] \;.
\end{equation}
Therefore Eq. (\ref{Pqu2}) reads for all time $t$
\begin{eqnarray} \label{def_phi}
\sum_{Q=0}^{\infty} e^{-pQ} \,P_{\rm qu} (Q,t) = \exp\left[-\rho \sqrt{D t}~ \phi (p)\right] \;.
\end{eqnarray}
In the long time limit $t \rightarrow \infty$, we anticipate, and verify a posteriori, that 
$P_{\rm qu}(Q,t)$ takes a large deviation form in the limit where $Q \to \infty$, $t \to \infty$ but
with the dimensionless ratio $q = Q/(\rho \sqrt{D\,t})$ fixed 
\begin{equation}\label{Pq}
P_{\rm qu}(Q,t) \sim \exp\left[-\rho \sqrt{D t}~ \Psi_{\rm diff} \left(\frac{Q}{\rho \sqrt{D t}}\right)\right], 
\end{equation}
where $\Psi_{\rm diff}(q)$ is a rate function that we wish to compute. Substituting this large deviation form (\ref{Pq}) on the left hand side (lhs) of Eq. (\ref{def_phi}) and replacing the discrete sum over $Q$ by an integral (which is valid for large $Q \sim \sqrt{t}$), we get
\begin{equation}\label{Pq1}
\int_0^{\infty} e^{-pQ} P_{\rm qu}(Q,t) dQ \sim \rho \sqrt{D t} \int_0^{\infty} e^{-\rho \sqrt{D t}\, \left[p \, q+\Psi_{\rm diff}(q)\right]} dq \;.
\end{equation}
For large $t$, we can now evaluate the integral over $q$ in Eq. (\ref{Pq1}) by a saddle point method, which gives
\begin{eqnarray}\label{sp_diff}
\int_0^{\infty} e^{-pQ} P_{\rm qu}(Q,t) dQ \sim \exp{\left[-\rho \sqrt{Dt}\, \underset{q}{\min}[p\,q+\Psi_{\rm diff}(q)] \right]} \;.
\end{eqnarray}
Comparing this with the rhs of Eq. (\ref{def_phi}) we get
\begin{equation}\label{legendre}
\underset{q}{\min}\left[p\,q+\Psi_{\rm diff}(q)\right]= \phi(p) \;.
\end{equation}
Inverting this Legendre transform one gets
\begin{equation}\label{fq-bas}
\Psi_{\rm diff}(q)=\underset{p}{\max}\left[\phi(p)-p\,q\right] \;,
\end{equation}
where $\phi(p)$ is given in Eq. (\ref{phi-1eq}). 

\begin{figure}[htb]
\begin{center}
\begin{minipage}{0.4\hsize}
\includegraphics[width=\hsize]{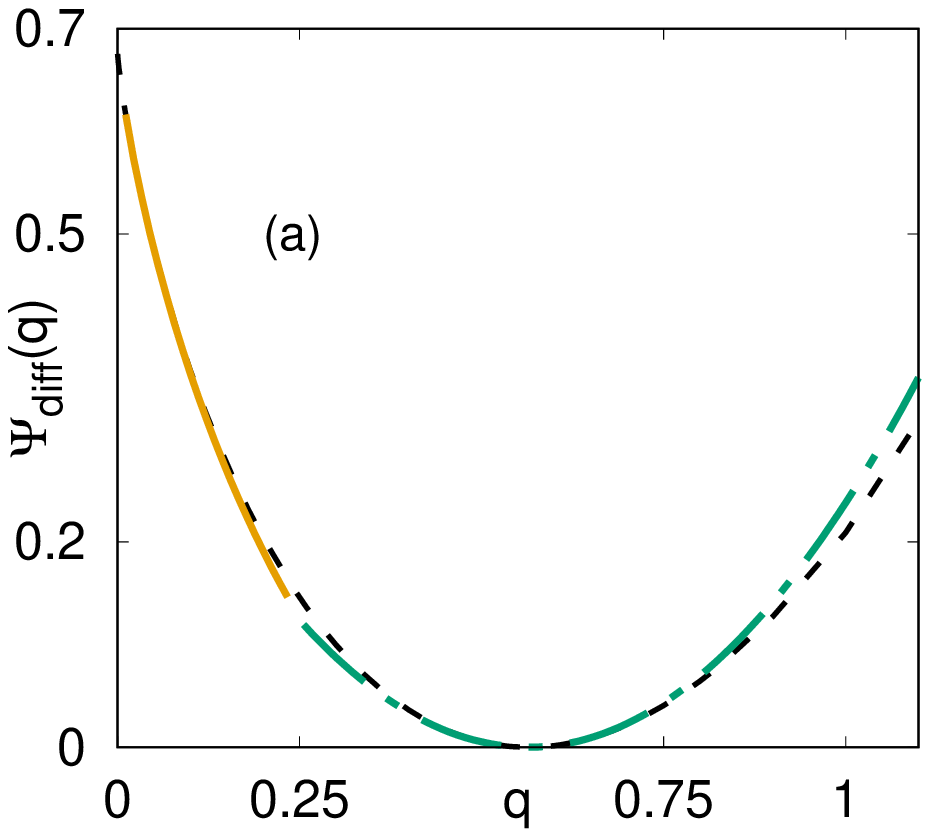}
\end{minipage}
\begin{minipage}{0.4\hsize}
\includegraphics[width=\hsize]{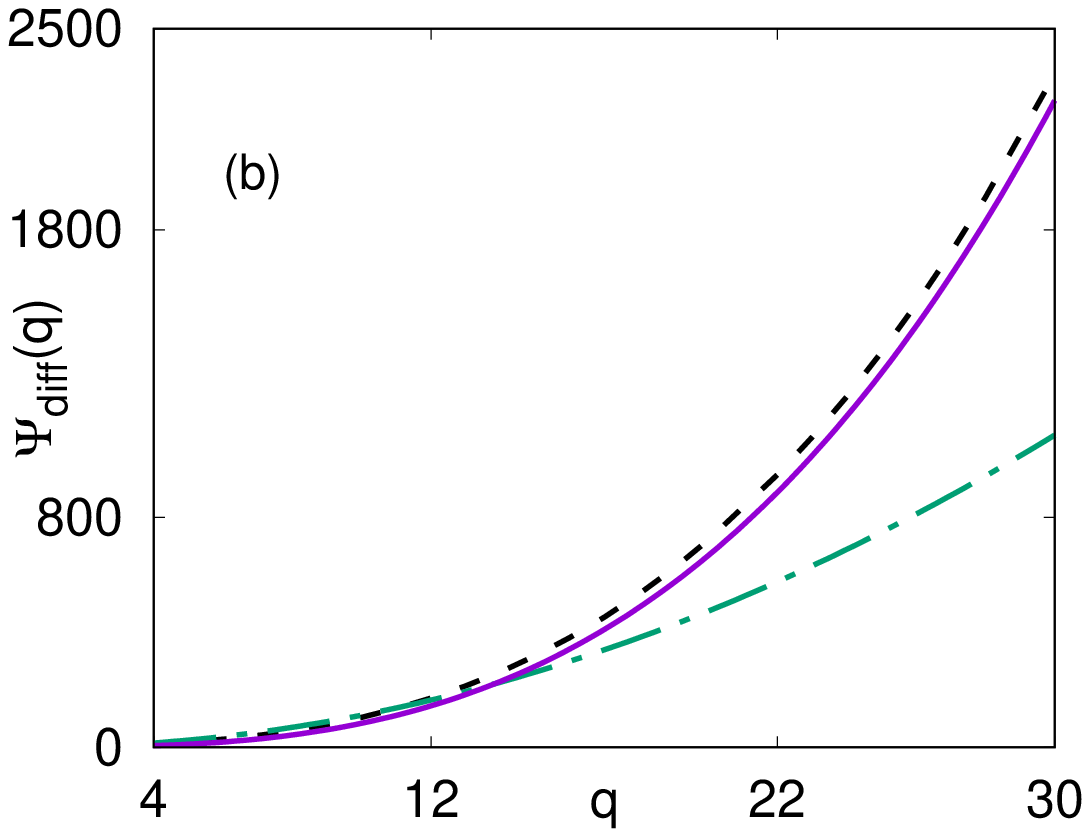}
\end{minipage}
\end{center}
\caption{Large deviation function $\Psi_{\rm diff}(q)$ vs $q$ for the diffusive case with quenched initial conditions. On both panels, the dashed black lines correspond to 
evaluating via Mathematica $\Psi_{\rm diff}(q)$ from Eq. (\ref{fq-bas}) with $\phi(p)$ given in Eq. (\ref{phi-1eq}). On the left panel (a), the solid yellow curve corresponds to the small $q$ asymptotic behavior of $\Psi_{\rm diff}(q)$ in Eq.~(\ref{Fq-qzero-soln}) and the dashed-dotted green curve corresponds to the quadratic behavior in Eq.~(\ref{fq_s0}). On the right panel (b), we zoom in on the large $q$ tail. The solid violet curve corresponds to the leading asymptotic behavior $\Psi_{\rm diff}(q) \approx q^3/12$, while the dashed-dotted green curve is the quadratic behavior as in Eq.~(\ref{fq_s0}). The violet and the green curves clearly demonstrate the non-Gaussian tail of $\Psi_{\rm diff}(q)$.}\label{diff-math-asymp}
\end{figure}

Knowing $\phi(p)$ explicitly, one can plot the large deviation function $\Psi_{\rm diff}(q)$ using Eq. (\ref{fq-bas}) -- see Fig. \ref{diff-math-asymp}. Clearly,
$\Psi_{\rm diff}(q)$ has a concave shape with a minimum at $q=q_{\min}$, the value of $q_{\min}$ will be computed shortly. The asymptotic behaviors of the
rate function $\Psi_{\rm diff}(q)$ can also be extracted in the limits $q \to q_{\min}$, $q\to 0$ and $q \to \infty$ by analysing $\phi(p)$ respectively in the limits $p\to 0$, $p\to +\infty$ and $p \to -\infty $ (where $\phi(p)$ in Eq. (\ref{phi-1eq}) has to be continued analytically to negative $p$). The results are summarised in Eqs. (\ref{psi_diff_asympt}), (\ref{C}) and (\ref{A}) in section \ref{sec:model}. Below we provide the derivation of these results.

\subsubsection{Typical Fluctuations : $Q \sim \langle Q \rangle_{\rm qu}$}\label{qmi}
 In order to derive the result for $\Psi_{\rm diff}(q \rightarrow q_{\min})$, we need to analyze $\phi(p)$ for $p \rightarrow 0$. We expand $\phi(p)$ in Eq. (\ref{phi-1eq}) up to order $p^2$ and get
\begin{eqnarray}\label{phi_s0}
\phi(p) = \alpha p - \beta p^2 + {\cal O}(p^3) \;, 
\end{eqnarray}
where 
\begin{eqnarray}
&&\alpha=\int_0^{\infty} {\rm erfc}(z) dz ~ = \sqrt{\frac{1}{\pi}} \label{alpha-defn} \\
&&\beta=\frac{1}{4} \int_0^{\infty} (2\, {\rm erfc}(z) - {\rm erfc}^2(z))\, dz ~ = \frac{1}{\sqrt{8\pi}}\;.
\end{eqnarray}
Substituting $\phi(p) = \alpha \, p - \beta\,p^2$ in Eq. (\ref{fq-bas}) and maximising with respect to $p$ gives a quadratic form for the rate function 
 \begin{equation} \label{fq_s0}
 \Psi_{\rm diff }(q) \sim \frac{(q-\alpha)^2}{4\beta}= \sqrt{\frac{\pi}{2}}     \left(q-\sqrt{\frac{1}{\pi}}\right)^2 \;.
 \end{equation}
This form holds for $q$ close to $q_{\min} = \alpha = 1/\sqrt{\pi}$ and gives the result in the second line in Eq. (\ref{psi_diff_asympt}). Substituting this quadratic behavior in the large deviation form in Eq. (\ref{Pq}) predicts
a Gaussian form for the quenched flux distribution for $q$ close to $q_{\min}$
\begin{eqnarray}\label{Gaussian_qu}
P_{\rm qu}(Q,t) \sim \exp{\left[- \frac{\left(Q-\langle Q\rangle_{\rm qu}\right)^2}{2 \sigma_{\rm qu}^2}\right]}
\end{eqnarray}
where the mean the variance are given by 
 \begin{eqnarray}
\langle Q \rangle_{\rm qu} &=&\rho \sqrt{\frac{D t}{\pi}}  \label{diffusion-mean-quenched} \\
\sigma^2_{\rm qu} &=& \rho \sqrt{\frac{D t}{2\pi}} \label{diffusion-variance-quenched} \;.
\end{eqnarray} 
Notice that these expressions for the mean and the variance, though derived here for large $t$, actually hold for all $t$, as one can verify directly from the formulae in Eqs. (\ref{mean_qu}) and (\ref{var_qu}) with $U(z,t) = (1/2) {\rm erfc}(z/\sqrt{4 Dt})$. Comparing with the annealed case, while the means in both cases are identical, both given by $\mu(t) = \rho \sqrt{Dt/\pi}$, their variances and higher moments differ. For example, the variance in the annealed case is $\mu(t)=\rho \sqrt{Dt/\pi}$ which differs by a factor $1/\sqrt{2}$ from the quenched case in Eq. (\ref{diffusion-variance-quenched}). These results agree with those obtained in~\cite{derrida-gers}. This typical quadratic behavior is shown by the dashed-dotted green curve in Fig. \ref{diff-math-asymp}.

\subsubsection{Atypical fluctuations on the left of the mean: $Q \ll \langle Q \rangle_{\rm qu}$} \label{qze}

In order to infer about the fluctuations of $P_{\rm qu}(Q,t)$ around $Q \rightarrow 0$, we need to evaluate how $\Psi_{\rm diff}(q)$ behaves when $q \rightarrow 0$. This corresponds to the limit $p \rightarrow \infty$ for $\phi(p)$ from Eq. (\ref{fq-bas}). We use the large $p$ expansion in Eq. \ref{Ip-pinf1}, and evaluate $A(t)$ and $B(t)$ from 
Eq. (\ref{atbt-pinf1}) using $U(z,t) = (1/2) {\rm erfc}(z/\sqrt{4Dt})$. This gives
\begin{eqnarray}\label{At-diff}
A(t) &=& -\overline{\alpha} \, \rho\, \sqrt{4Dt} \;, \quad {\rm where} \quad \overline{\alpha}= -2 \int_0^{\infty} {\ln} \left[1-\frac{1}{2}{{\rm erfc}(z)}\right] dz =  0.675336\ldots  \label{a-defn} \\ 
B(t) &=&  \overline{\beta} \, \rho\, \sqrt{4Dt} \;, \quad {\rm where} \quad \overline{\beta} = \int_0^{\infty} \frac{{\rm erfc}(z)}{1-\frac{1}{2}{\rm erfc}(z)} dz = 0.828582 \ldots \;. \label{b-defn}
\end{eqnarray}
From Eqs. (\ref{I-phi-correspondence}) and (\ref{phi-1eq}) we get the two leading terms of $\phi(p)$ for large $p>0$
\begin{equation}\label{phip-zero-final}
\phi(p)= -\frac{I(p,t)}{\rho \sqrt{D t}} \approx \overline{\alpha}-\overline{\beta}~e^{-p}  \;.
\end{equation}
Plugging this result for $\phi(p)$ in Eq. (\ref{fq-bas}) and maximizing with respect to $p$, we get the leading small $q$ behavior of $\Psi_{\rm diff}(q)$
\begin{equation}\label{Fq-qzero-soln}
\Psi_{\rm diff}(q)\approx \overline{\alpha} - q +q ~ {\ln}\left(\frac{q}{\overline{\beta}}\right) \;.
\end{equation}
This reproduces the first line of Eq. (\ref{psi_diff_asympt}). In particular, for $q=0$, using $\Psi_{\rm diff}(q=0) = \overline{\alpha}$ in Eq. (\ref{Pq}), we obtain $P_{\rm qu}(Q=0,t) \sim \exp{(-\overline{\alpha} \, \rho \sqrt{Dt})}$ as announced in Eq. (\ref{P_Q0}). The small $q$ behavior of $\Psi_{\rm diff}(q)$ is shown by the solid yellow curve in  Fig. \ref{diff-math-asymp}(a).

\subsubsection{Atypical fluctuations on the right of the mean: $Q \gg \langle Q\rangle_{\rm qu}$}\label{qla}

In order to derive the large $q$ asymptotics of $\Psi_{\rm diff}(q)$ from Eq. (\ref{fq-bas}), we first need to continue $\phi(p)$ in Eq. (\ref{phi-1eq}) analytically to negative $p$
and use its asymptotics in the limit $p \to -\infty$. For this, it is convenient to write first $p = -u$ where $u = |p|$. We write
\begin{equation}\label{tildephi-diff}
\phi(p=-u) = \tilde{\phi}(u) = -2 \int_0^{\infty} dz ~{\ln} \left[1+ \frac{\left(e^u-1 \right)}{2}{{\rm erfc}(z)} \right] \quad \underset{u \to \infty}{\approx} \quad -2 \int_0^{\infty} dz ~{\ln} \left[1+ \frac{e^u}{2} {\rm erfc}(z) \right] \;.
\end{equation}
To extract the large $u$ behavior of $\tilde \phi(u)$ from the integral on the rhs, it is convenient to take the derivative with respect to $u$
\begin{equation}\label{qu-larphi}
 \tilde{\phi}'(u) \approx -2\int_0^{\infty} dz \, \frac{~ \frac{e^{u}}{2}\,{\rm erfc}(z)}{1 + \frac{e^u}{2}\,{\rm erfc}(z)} \;.
\end{equation} 
For large $u$ the dominant contribution to this integral comes from large $z$ where ${\rm erfc}(z) \approx e^{-z^2}/(z \sqrt{\pi})$. Hence we see that, for $z > \sqrt{u}$ the integrand is essentially $0$ as $u \to \infty$, while, for $z < \sqrt{u}$, the integrand is $1$ as $u \to \infty$. Hence, the integrand can be approximated by a Fermi function 
\begin{eqnarray} \label{phiprime2}
\tilde{\phi}'(u) \approx -2 \int_0^{\sqrt{u}} dz = - 2 \sqrt{u} \;.
\end{eqnarray}
Integrating it back, we get the leading order behavior for $\tilde \phi(u)$ for large $u$
\begin{equation}\label{phi-large-diff}
 \tilde{\phi}(u) \approx  -\frac{4}{3} u^{\frac{3}{2}} \;.
\end{equation}
Therefore $\phi(p) \approx -(4/3) (-p)^{3/2}$ as $p \to -\infty$. Substituting this behavior in Eq. (\ref{fq-bas}) and maximizing with respect to $p$ one gets $\Psi_{\rm diff}(q) \approx q^3/12$ as $q \to \infty$. This then gives the last line of the result in Eq. (\ref{psi_diff_asympt}). As mentioned earlier, this leading large $q$ asymptotic behavior of $\Psi_{\rm diff}(q)$ coincides with the result of Ref. \cite{derrida-gers} obtained by a different method. The large $q$ behavior of $\Psi_{\rm diff}(q)$ is shown solid the solid violet curve in Fig. \ref{diff-math-asymp}(b).

\subsection{$P_{\rm qu}(Q,t)$ for run-and-tumble particles}\label{quen-act-sec}

In this case, our starting point again are Eqs. (\ref{Pqu2}) and (\ref{def_Ip}), except that the function $U(z,t)$ for the RTP is more complicated.
Its Laplace transform is given in Eq. (\ref{LaplaceU_2}). As shown in Appendix C of \cite{pierre-satya-greg} it can be formally inverted to obtain $U(z,t)$ in real time
\begin{equation}\label{Uzt-exact}
U(z,t) = \frac{1}{2}\left[ e^{\frac{-\gamma z}{v_0}} + \frac{\gamma z}{v_0} \int_1^{\frac{v_0 t}{z}} dT \frac{ e^{\frac{-\gamma z T}{v_0}}I_1(\frac{\gamma z}{v_0}\sqrt{T^2-1})}{\sqrt{T^2-1}}\right] \Theta(v_0 t-z)  \;.
\end{equation}
However, it turns out that this expression is not very useful to extract the large deviation function at late times. 

Before proceeding to compute the large deviation function at late times, it is useful to discuss the large $t$ behavior of $P_{\rm qu}(Q,t)$ in different
regimes of $Q$. In the following, we will first discuss the $Q \to 0$ limit of $P_{\rm qu}(Q,t)$, followed by the discussion of the typical
fluctuations where $Q = {\cal O}(\sqrt{t})$. In this regime, we will recover the Gaussian fluctuations. When $Q/(\rho\, \sqrt{D\,t}) \gg 1$, we
expect to recover the large deviation regime for the diffusive behaviour discussed in the previous section. This is because, as explained in the 
introduction, at late times, the RTP motion essentially reduces to that of a diffusive particle with an effective diffusion constant $D_{\rm eff} = v_0^2/(2 \gamma)$.
However, there exists yet another ``larger deviations regime'' where $Q \sim {\cal O}(t)$ where we will show that $P_{\rm qu}(Q,t)$ carries the signature
of activity and has a novel large deviation form
\begin{equation}\label{ac-sig2}
P_{\rm qu}(Q,t) \sim \exp\left[-\rho \, v_0 \, \gamma \, t^2 \, \Psi_{\rm RTP}\left(\frac{Q}{\rho \, v_0 \,t}\right)\right] \;.
\end{equation}
In the following, we will indeed compute this rate function $\Psi_{\rm RTP}(q)$ and show that it is given by Eq. (\ref{rtp-ldf-model}).

\subsubsection{Typical fluctuations: $Q \sim \langle Q\rangle_{\rm qu}$}

In order to extract the typical fluctuations of $Q_t$ around its mean value for the RTP case, we need to use the small $p$ expansion
of $I(p,t)$ in Eq. (\ref{Pqu2}). Quite generally, the small $p$ expansion of $I(p,t)$ is given in Eq. (\ref{Ip0}). We use this expansion
on the rhs of Eq. (\ref{Pqu2}) and approximate the sum on lhs by an integral. The resulting Laplace transform can be easily inverted
and yields a Gaussian form 
\begin{eqnarray}\label{P_qu_Gauss}
P_{\rm qu}(Q,t) \approx \exp{\left[-\frac{(Q-\langle Q \rangle_{\rm qu})^2}{2 \, \sigma_{\rm qu}^2}\right]} \;,
\end{eqnarray} 
where $\langle Q \rangle_{\rm qu}$ and $\sigma^2_{\rm qu}$ are given in Eqs. (\ref{mean_qu}) and (\ref{var_qu}) respectively where $U(z,t)$ is given in Eq. (\ref{Uzt-exact}) --
alternatively its Laplace transform is given by the simpler form in Eq. (\ref{LaplaceU_2}). The mean value $\langle Q \rangle_{\rm qu}$ can be computed explicitly. Indeed
\begin{eqnarray}\label{av_qu_RTP}
\langle Q \rangle_{\rm qu} = \rho \int_0^\infty U(z,t)\, dz = \mu(t)=  \frac{\rho\,v_0}{2} t\,e^{-\gamma t}\left[ I_0(\gamma\,t) + I_1(\gamma \, t)\right] \;,
\end{eqnarray}
where the last equality follows from Eq. (\ref{ac-mut}). The variance $\sigma_{\rm qu}^2 =  \rho \int_0^{\infty} U(z,t)\left[1-U(z,t)\right] \, dz$ is however difficult to compute explicitly using $U(z,t)$ from Eq. (\ref{Uzt-exact}). However, it can be easily evaluated numerically. At large times, $\langle Q \rangle_{\rm qu}$ and $\sigma^2_{\rm qu}$ converge to the diffusive limits given in Eqs. (\ref{diffusion-mean-quenched}) and (\ref{diffusion-variance-quenched}) respectively.

\subsubsection{Atypical fluctuations on the left of the mean: $Q \ll \langle Q \rangle_{\rm qu}$}

Exactly at $Q=0$ or $Q=1$, we have an exact
expression at all times $t$ for $P_{\rm qu}(Q,t)$ in terms of $U(z,t)$, as given in Eqs. (\ref{Pqu_eq_0})
and (\ref{Pqu_eq_1}). The function $U(z,t)$ for RTP appearing in these expressions is given in Eq. (\ref{Uzt-exact}).
Given this rather complicated expression of $U(z,t)$, it is hard to obtain explicit formulae valid at all times for
$P_{\rm qu}(Q,t)$ even for $Q=0$ or $Q=1$. However, at late times, since $U(z,t)$ converges at late times
to that of the diffusive limit in Eq. (\ref{UzT_BM}) with an effective diffusion constant $D_{\rm eff} = v_0^2/(2 \gamma)$, 
we recover the diffusive results for this extreme left tail of $P_{\rm qu}(Q,t)$.
For instance $P_{\rm qu}(Q=0,t)$, which represents the probability of having no particle on the right side of the origin
at time $t$, decays at late times as in the diffusive case 
\begin{eqnarray} \label{P_Q0_RTP}
P_{\rm qu}(Q=0,t)\Big \vert_{\rm RTP} \approx \exp \left[ - \overline{\alpha} \, \rho \sqrt{D_{\rm eff}\,t}\right] \;,
\end{eqnarray} 
where $\overline{\alpha} = 0.675336\ldots$ is given in Eq. (\ref{C}).

\subsubsection{Atypical fluctuations on the right of the mean: $Q \sim {\cal O}(t) \gg \langle Q \rangle_{\rm qu}$}

In this section, we derive the result in Eq. (\ref{ac-sig2}). We recall that in the diffusive case, the atypical 
fluctuations of $Q$ are encoded in the large deviation form in Eq. (\ref{Pq}) with $Q \sim \rho \sqrt{Dt}$. 
The extreme fluctuations to the right of $\langle Q \rangle_{\rm qu}$ in this case are described by the large
argument behavior of the large deviation function ${\Psi}_{\rm diff}(q = Q/(\rho\sqrt{D\,t}))$, i.e., when $Q \gg \rho \sqrt{Dt}$.   
Thus, in the diffusive case, there is a single scale for the fluctuations of $Q$ at late times, namely $Q \sim \sqrt{t}$. 
In contrast, for the RTP, in addition to the scale $\sqrt{t}$ that describes the moderate large deviations around
the mean, there is yet another scale where $Q \sim t$. This comes from the fact that each particle in time $t$ can move a maximum distance 
$v_0 t$, where $v_0$ is the velocity. So for an initial density $\rho$, the maximum possible flux through the origin is $Q_{\rm max} = \rho v_0 t$. 
Hence $Q \sim t$ describes the scale of fluctuations at the very right tail of the distribution $P_{\rm qu}(Q,t)$.

To extract this extreme right tail, we again start from Eqs. (\ref{Pqu2}) and (\ref{def_Ip}) with $U(z,t)$, for RTP, given by its Laplace transform 
in Eq. (\ref{LaplaceU_2}). Before extracting the large deviation form of $P_{\rm qu}(Q,t)$, we first analyse $U(z,t)$ in the limit $z \sim t$. Inverting the Laplace transform of $U(z,t)$ in Eq. (\ref{LaplaceU_2}) we get
\begin{eqnarray}\label{U_Bromwich1}
U(z,t) = \int_\Gamma \frac{ds}{2\pi i} \exp{\left[t\left( s - \sqrt{s(s+2\gamma)} \frac{z}{v_0\,t}\right)\right]} \;,
\end{eqnarray}   
where $\Gamma$ represents the Bromwich contour in the complex $s$-plane. In the limit $t \to \infty$, $z \to \infty$ with the ratio $z/t$ fixed,
the integral can be evaluated by the saddle-point method, which yields (up to pre-exponential factors)
\begin{equation}\label{Uzt_saddle}
U(z,t) \approx \exp \left[-\gamma t \left(1-\sqrt{1-\frac{z^2}{v_0^2 t^2}}\right)\right] \, \Theta(v_0 \,t-z ) ~.
\end{equation}
We have checked numerically that this approximation (\ref{Uzt_saddle}) works very well, at large times, as one would expect. 

To extract the large $Q \sim t \gg \langle Q\rangle_{\rm qu}$ behavior from Eq. (\ref{Pqu2}) we need to analytically continue
$I(p,t)$ to $p$ negative and study the limit $p \to -\infty$, as in the diffusive case. Setting $p=-u$ with $u>0$, and approximating the discrete sum
on the lhs of Eq. (\ref{Pqu2}) by an integral, we get
\begin{eqnarray}\label{laplace1}
\int_0^\infty P_{\rm qu}(Q,t)\, e^{u Q} \,dQ \approx e^{\tilde I(u,t)} \;,
\end{eqnarray}
where
\begin{eqnarray}\label{def_tildeI}
\tilde I(u,t) = \rho \int_0^\infty dz\, \ln \left( 1+ (e^u-1)\, U(z,t)\right) \underset{u \to \infty}{\approx}   \rho \int_0^\infty dz\, \ln \left( 1+ e^u\, U(z,t)\right) \;.
\end{eqnarray}
To extract the large $u$ behavior of $\tilde I(u,t)$ we follow the same procedure as in the diffusive case and take a derivative with respect to $u$. 
We get
\begin{eqnarray}\label{derivative_1}
\frac{d\tilde I(u,t)}{du} \approx \rho \int_0^\infty \frac{dz}{1+e^{-u} \frac{1}{U(z,t)}} \;.
\end{eqnarray}
For large $u$, this integral is dominated by the region where $z \sim t$ where we can use the approximate form of $U(z,t)$ given in Eq. (\ref{Uzt_saddle}). Substituting this form for $U(z,t)$ in Eq. (\ref{derivative_1}), we get
\begin{eqnarray}\label{derivative_2}
\frac{d\tilde I(u,t)}{du} \approx \rho \int_0^{v_0\,t} \frac{dz}{1+\exp{\left[-\left(u-\gamma\,t  + \gamma\,t \sqrt{1-\frac{z^2}{v_0^2\,t^2}}\right)\right]}} \;.
\end{eqnarray}
We now analyse this integral in two different cases, assuming $u \sim t \gg 1$:

\vspace*{0.3cm}
\noindent $\bullet$ If $u > \gamma t$: in this case as $t \to \infty$ it is clear that the integrand in Eq. (\ref{derivative_2}) is always $1$ for any $z$. Hence 
\begin{eqnarray}\label{derivative_3}
\frac{d\tilde I(u,t)}{du} \approx \rho v_0 t \quad \quad {\rm if} \quad u > \gamma t \;.
\end{eqnarray}

 \begin{figure}
\begin{center}
\begin{minipage}{0.32\hsize}
\includegraphics[width=\hsize]{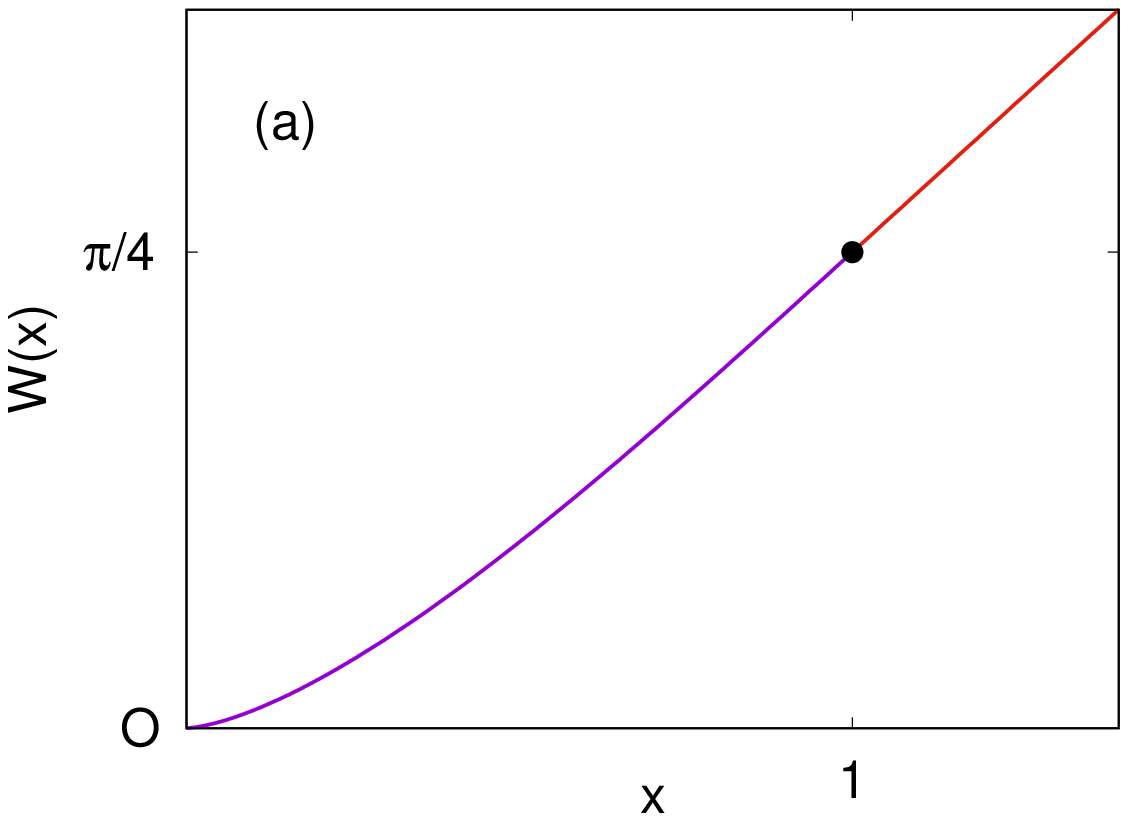}
\end{minipage}
\begin{minipage}{0.32\hsize}
\includegraphics[width=\hsize]{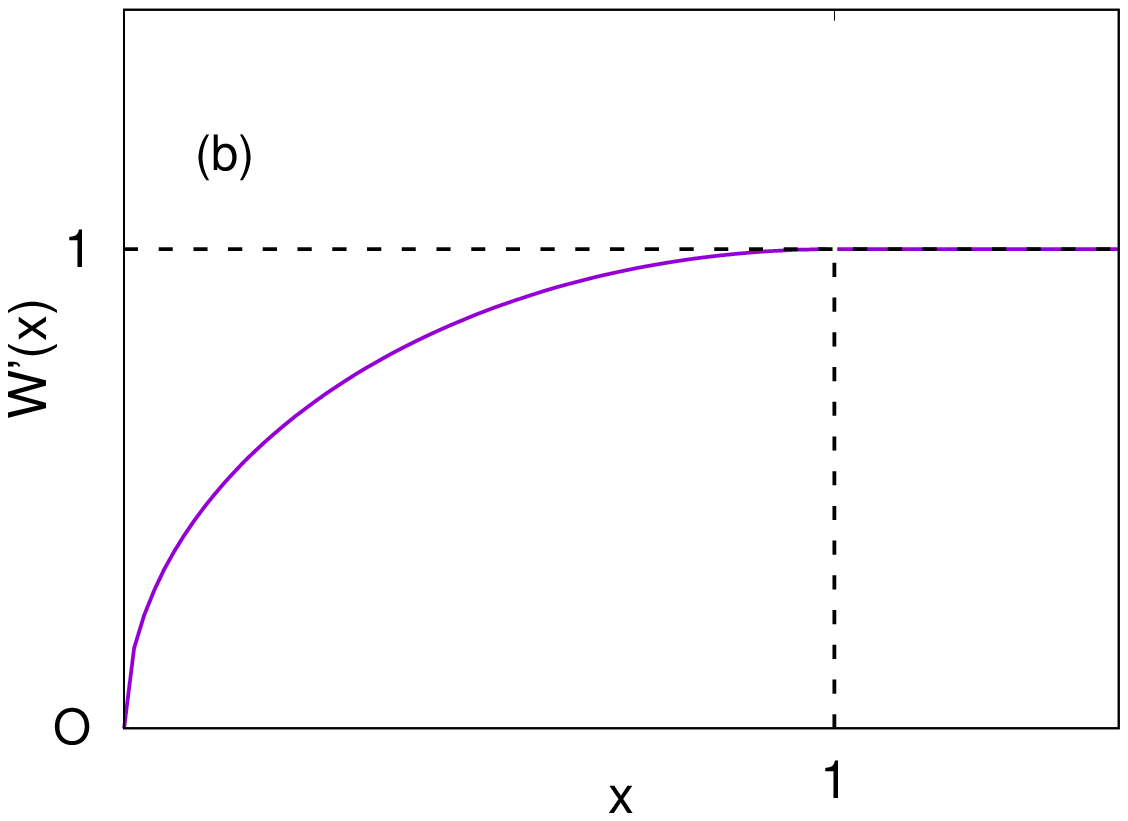}
\end{minipage}
\begin{minipage}{0.32\hsize}
\includegraphics[width=\hsize]{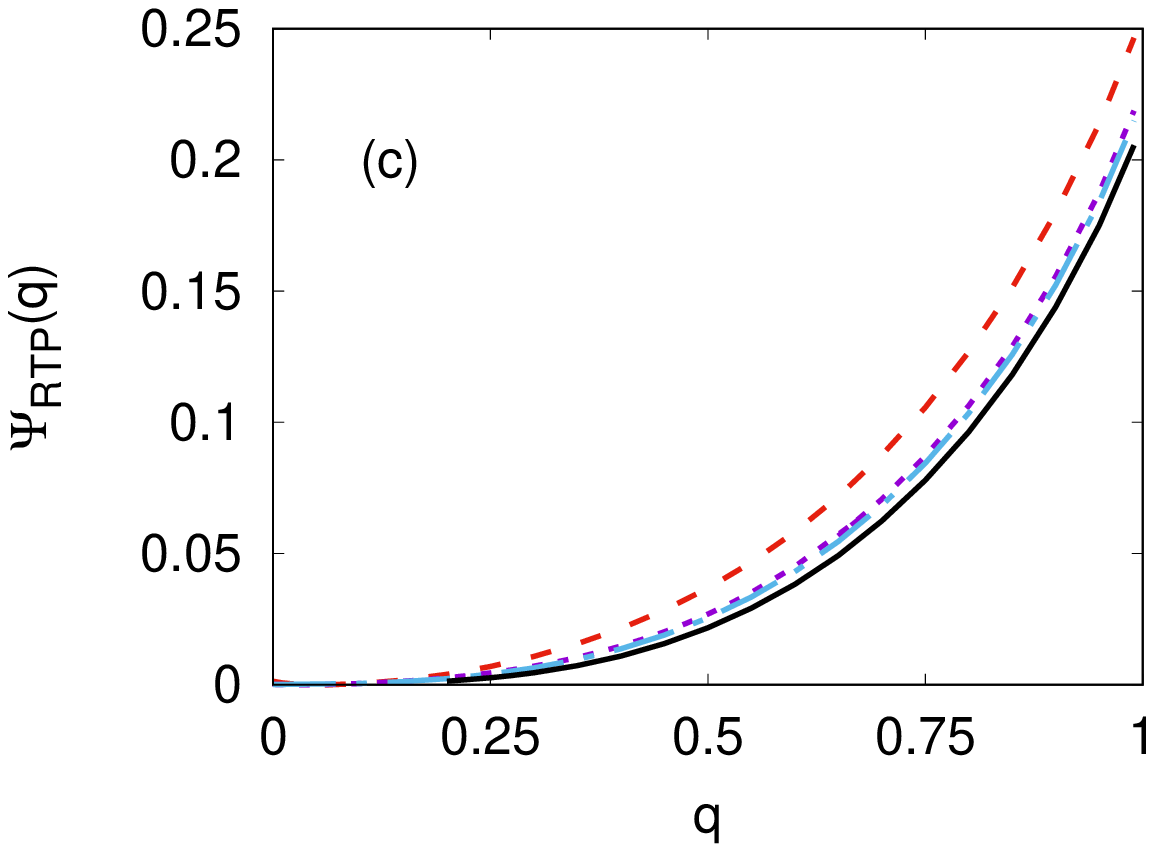}
\end{minipage}
\end{center}
\caption{ (a) Plot of $W(x)$ vs $x$ with $W(x)$ given by Eq.~(\ref{phi-full}). The non-analytical point at $x=1$ is shown by a black dot
where $W(x)$ and its first two derivatives are continuous, while the third derivative is discontinuous, as in Eq. (\ref{third-der-phi}). 
(b) Plot of $W'(x)$ vs $x$ where $W'(x)$ can be read off from Eq. (\ref{def_W}). Note that $W'(x) <1$ for $x<1$ and saturates at $1$ at $x=1$. (c) Large deviation function $\psi_{\rm {RTP}}(q)$ vs $q$ for the run-and-tumble case. Black solid line represents the analytical expression given in Eq.~(\ref{rtp-ldf-model}) valid at very large time. Dashed lines ($t=100$ (red-dashed), $t=400$ (violet-dotted) and $t=600$ (blue dashed-dotted)) correspond to the finite time evaluation of $\underset{x}{\max}[\frac{qu}{\gamma t}-\frac{\tilde{I}(u,t)}{\rho v_0 \gamma t^2}]$ using Eq. (\ref{def_tildeI}) -- 
together with  Eq.~(\ref{def_tildeI}) with $\rho=v_0=1, \gamma=0.5$ -- with Mathematica.
}\label{phi-full-plot}
\end{figure}

\vspace*{0.3cm}
\noindent $\bullet$ If $u < \gamma t$: this case is a bit more complicated to analyze. Since $u<\gamma t$ the argument of the exponential in Eq. (\ref{derivative_2}) can be either positive or negative. Accordingly, the integrand will either $0$ or $1$ for large $u \sim t \gg 1$. The value
of $z$ for which the argument of the exponential changes sign is given by
\begin{eqnarray}\label{zstar}
z^*(u) = \frac{v_0}{\gamma} \sqrt{u(2\gamma\,t - u)} \;.
\end{eqnarray}
Thus, if $z>z^*(u)$ the integrand is $0$ while for $z<z^*(u)$ the integrand is $1$. Hence this integral over $z$ in (\ref{derivative_2}) gets cut-off at
$z=z^*(u)$ (note that $z^*(u)<v_0\,t$ for all $u$). Hence we get
\begin{eqnarray}\label{derivative_Iu}
\frac{d\tilde I(u,t)}{du} \approx \rho \, z^*(u) = \rho \frac{v_0}{\gamma} \sqrt{u(2 \gamma t-u)} \quad \quad {\rm if} \quad u < \gamma \,t \;.
\end{eqnarray} 
Integrating it back with respect to $u$ we obtain  
\begin{eqnarray}\label{full-intI_1}
\tilde{I}(u,t) \approx \rho v_0 \gamma \, t^2 \, W\left(\frac{u}{\gamma t} \right) 
\end{eqnarray}
where 
\begin{eqnarray}\label{def_W}
W(x) =  
\begin{cases}
\int_0^{x} \sqrt{y(2-y)}dy\quad,\quad ~~~~~~~\;\;\;\;\;\;{\rm if}~ x <1 \;,\\ 
\\
\int_0^1 \sqrt{y(2-y)}dy+ (x-1) \quad,\quad {\rm if} ~ x>1 \;.
\end{cases}
\end{eqnarray}
Performing these integrals explicitly, we get
\begin{eqnarray}\label{phi-full}
W(x) =
\begin{cases}
 \frac{(x-1)}{2}\sqrt{x(2-x)}+{\rm sin}^{-1}(\sqrt{x/2})\quad,\quad x<1\\ 
\\
\frac{\pi}{4}+x-1\quad,\quad x>1 \;.
\end{cases}
\end{eqnarray}
The function $W(x)$ is plotted vs $x$ in Fig.~\ref{phi-full-plot} (a). Interestingly, while $W(x)$ and its first two derivatives are continuous 
at $x=1$, its third derivative is discontinuous. Indeed one has
\begin{eqnarray}\label{third-der-phi}
W^{'''}(x \rightarrow 1^-) &=& -1 \nonumber \\
W^{'''}(x \rightarrow 1^+) &=& 0 \;.
\end{eqnarray}
Using this result for $W(x)$ in Eq. (\ref{phi-full}) gives us an expression for $\tilde I(u,t)$ in (\ref{full-intI_1}). We then substitute this expression for $\tilde I(u,t)$ in Eq. (\ref{laplace1}) to get
\begin{eqnarray}\label{PqW}
\int_0^\infty P_{\rm qu}(Q,t) e^{uQ} dQ  \sim \exp{\left[\rho v_0 \gamma t^2 W\left(\frac{u}{\gamma t}\right)\right]} \;.
\end{eqnarray}
Inverting formally this Laplace transform, we obtain
\begin{eqnarray} \label{Inverse1}
P_{\rm qu}(Q,t) \sim \int_{\Gamma} \frac{du}{2 \pi i} \exp{\left[- u Q + \rho v_0 \, \gamma t^2 W\left( \frac{u}{\gamma t}\right) \right]} \;.
\end{eqnarray}
Rescaling $u/(\gamma t)= x$ we get, up to pre-exponential factors,
\begin{eqnarray} \label{Inverse2}
P_{\rm qu}(Q,t) \sim \int_{\Gamma} \frac{dx}{2\pi i} \exp{\left[ - \rho v_0 \gamma t^2 \left(-W(x) + x q \right) \right]} \;, \quad {\rm where} \; q = \frac{Q}{\rho v_0 t} \;.
\end{eqnarray}
where $\Gamma$ is the Bromwich contour in the complex $x$-plane. Performing this integral by using a saddle-point for large $t$, we get
\begin{eqnarray} \label{Inverse3}
P_{\rm qu}(Q,t) \sim \exp{\left[- \rho v_0 \gamma t^2 \Psi_{\rm RTP} \left( q = \frac{Q}{\rho v_0 t}\right) \right]} \;,
\end{eqnarray}
with the rate function given by
\begin{eqnarray}\label{Psi_RTP}
\Psi_{\rm RTP}(q) = \underset{x}{\max} \left[q\,x - W(x) \right] \;,
\end{eqnarray}
where $W(x)$ is given explicitly in Eq. (\ref{phi-full}). It is easy to verify that the maximum of the function $q\,x-W(x)$ occurs at $x=x^* = 1-\sqrt{1-q^2} < 1$. Since $x^*<1$ we use the branch of $W(x)$ in the first line of Eq. (\ref{phi-full}). Substituting this value of $x^*$ in Eq. (\ref{Psi_RTP}) we get the result in Eq. (\ref{rtp-ldf-model}). The asymptotic behaviours of this function $\Psi_{\rm RTP}(q)$ are given in Eq. (\ref{asympt_RTP}) and a plot of this function is shown in Fig. \ref{phi-full-plot} (c). Note that for small $q$, i.e. $Q \ll \rho v_0 t$, $\Psi_{\rm RTP}(q)$ behaves as $\Psi_{\rm RTP}(q) \sim q^3/6$. Substituting this behavior in Eq. (\ref{Inverse3}) gives
\begin{eqnarray}\label{matching1}
P_{\rm qu}(Q,t) \Big \vert_{\rm RTP} \sim \exp{\left( - \frac{\gamma Q^3}{6 \rho^2 v_0^2 t}\right)} \sim \exp{\left( - \frac{Q^3}{12 \rho^2 D_{\rm eff}t}\right)} \;, \quad {\rm where} \quad D_{\rm eff} = \frac{v_0^2}{2 \gamma} \;.
\end{eqnarray}
On the other hand, for the diffusive case, from Eq. (\ref{P-F-diff-largeQ}), using $\Psi_{\rm diff}(q) \approx q^3/12$ for large $q$, i.e. $Q \gg \rho \sqrt{D\,t}$, one gets 
\begin{eqnarray}\label{matching2}
P_{\rm qu}(Q,t) \Big \vert_{\rm diff} \sim \exp{\left( - \frac{Q^3}{12 \rho^2 D_{}t}\right)}
\end{eqnarray}
Comparing these two tails (\ref{matching1}) and (\ref{matching2}), one sees that for the RTP, on a scale where $\rho \, \sqrt{D_{\rm eff}t} \ll Q \ll \rho \, v_0 \, t$ these two behaviors match perfectly, supporting the expectation that, at large $t$, even for moderately large fluctuations to the right of the mean, the flux distribution for the RTP and the diffusive case coincide, once one identifies the effective diffusion constant as $D_{\rm eff} = v_0^2/(2\gamma)$.

\section{Numerical Results}\label{numerics}

This section is dedicated to Monte Carlo simulations. We check numerically the analytical results  previously obtained and  characterize the properties of the physical realizations corresponding to large $Q$ values. 

\subsubsection{The Importance Sampling strategy}\label{IS-strategy}

 \noindent In order to obtain  the tails of $P_{an}(Q,t)$ and $P_{qu}(Q,t)$ we have employed Importance Sampling, a general method used to reduce the variance of observables whose expectation value is dominated by rare realizations, in this case rare trajectories. In the context of the evaluation of large deviation functions, very popular implementations of importance sampling ideas  are cloning algorithms  or transition path sampling \cite{giardina2011simulating}.
 Here we use an implementation of Importance Sampling, similar to transition path sampling, the details of the technique can be found in \cite{hartmann-epjb-2011, alberto-importance}.
Basically we sample realizations  with an exponential
bias  on their flux, $e^{-\theta Q}$. The adjustable parameter  $\theta$  allows to explore atypical realizations:
 a negative  $\theta$ favours realizations with large $Q$, while a positive  $\theta$ favours  small $Q$.  The sampling  is done using a standard Metropolis algorithm as discussed in \cite{hartmann-epjb-2011, alberto-importance} and error bars are smaller than the symbol size.

To proceed we note that the flux depends only  on  the  particle positions at time $t$ :
\begin{equation}\label{dis-diff}
 x_i(t) = x_i(0) +  \Delta x_i(t), \quad \forall i \;.
\end{equation}
We wrote this quantity as  the sum of two contributions: (i) the initial (negative) position,
 $x_i(0) <0$ and (ii) the total displacement, $\Delta x_i(t)$. The latter depends on the stochastic process we are considerning: for the diffusive particles it  is a Gaussian number of zero mean and standard deviation $\sqrt{2 D t}$.
For  active particles, it can be expressed as 
\begin{equation}\label{rtp-pos-velo}
 \Delta x_i(t) =\pm v_0 (T_1-T_2),
\end{equation}
where $T_1$,  is the total time spent moving in the  initial particle direction, $T_2=t-T_1$ the  time  spent in the opposite direction.
The signs $+$ or $-$ correspond to the initial direction and they are chosen  with equal probability. The times  $T_1$ and $T_2$ are determined as follows: the run times  $\tau_1, \tau_2,\ldots, \tau_n$ are drawn from an exponential distribution of rate $\gamma$, the last run being the first time  interval for which $\sum_{i=1}^{n} \tau_i > t$. Then $\tau_n$ is replaced by $t-\sum_{i=1}^{n-1} \tau_i$ and $T_1, T_2$ are computed.

The choice of the  initial conditions is the delicate point that makes the annealed case different from the quenched one: in the annealed case, averages are performed over all initial conditions while in the quenched case the initial condition is fixed and typical. We first study the annealed case for active particles. At large times, their behavior is statistically equivalent to the one of diffusive particles with the effective  constant $D_{\text{eff}}=v_0^2/(2 \gamma)$. Then we discuss the quenched case and recover the exact results of previous sections. At variance with the annealed case one expects that when $Q \simeq \rho v_0 t $ active particles should have a clear non-diffusive nature even at long times. Unfortunately, the biased Monte Carlo used in this paper does not allow to sample configurations with $Q \simeq \rho v_0 t$.

\begin{figure}
\begin{center}
\begin{minipage}{0.32\hsize}
\includegraphics[width=\hsize]{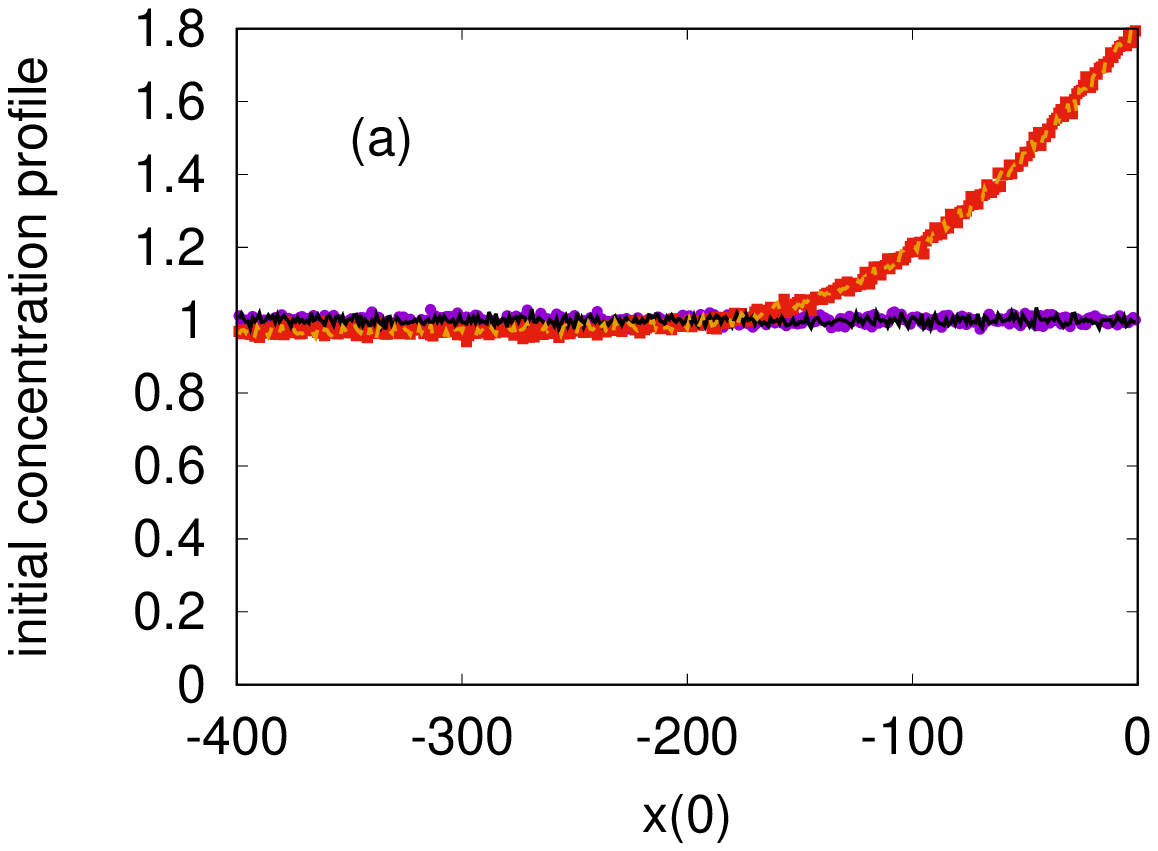}
\end{minipage}
\begin{minipage}{0.32\hsize}
\includegraphics[width=\hsize]{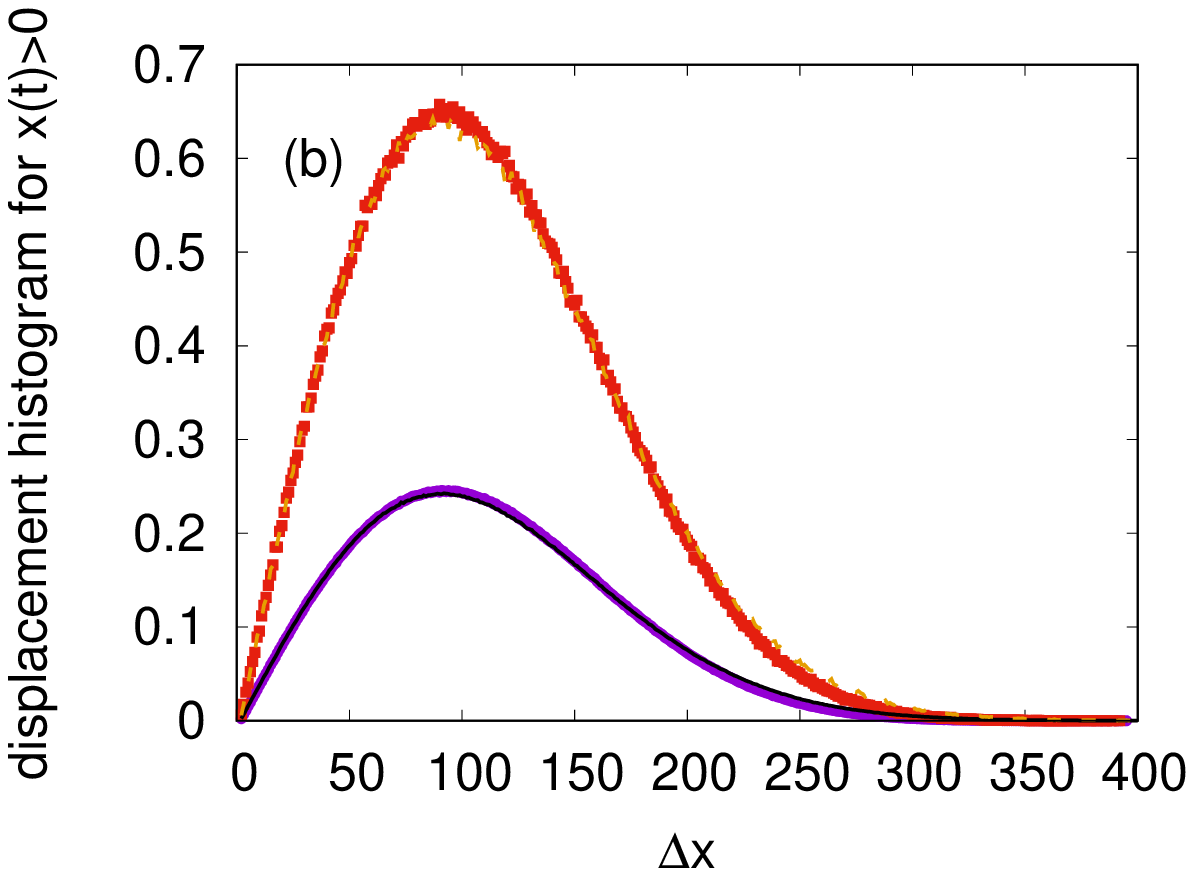}
\end{minipage}
\begin{minipage}{0.32\hsize}
\includegraphics[width=\hsize]{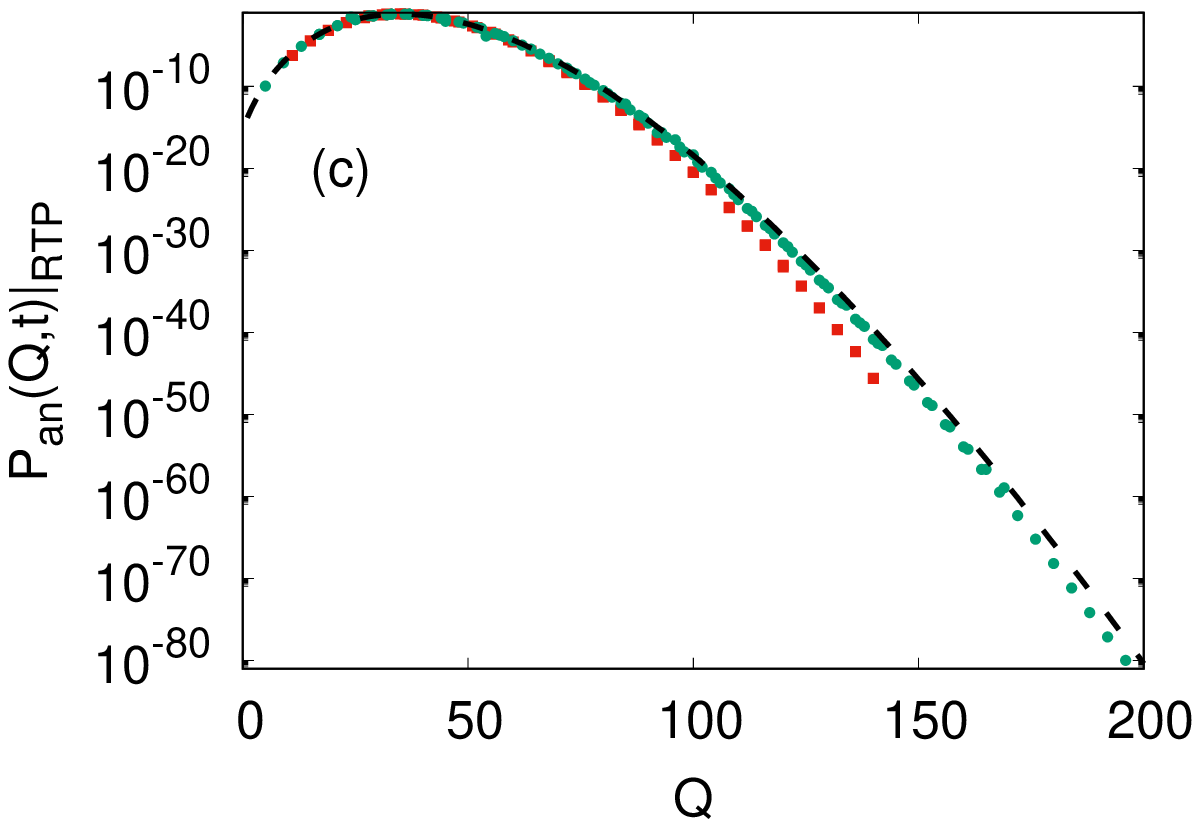}
\end{minipage}
\end{center}
\caption{Monte Carlo results for  the annealed case.  RTP  versus Diffusion. The initial particle position is sampled with $\rho=1$ and $L=2000$. Particles evolve up to time  $t=400$ with $v_0=1, \gamma=0.05$ for RTP (points), $D_{\text{eff}}=10$ for diffusion (lines). We performed  $10^7$ realizations.
 (a) Initial particle concentration.  For typical value of $Q$ (violet circles/black solid lines) the profile is flat, as expected. It has been  obtained by setting $\theta =0$ as importance sampling parameter  which corresponds to $Q \simeq  \langle Q \rangle \approx 35.45$. For large values of $Q$ (red squares/yellow dashed lines)   we observe an accumulation of particle close to origin. There we set $\theta =-1$ as importance sampling parameter which corresponds to  $ Q \approx 93 \gg \langle Q \rangle$.
(b) Histogram  (normalized to the number of realizations) of the displacement for particles with a positive final position with $\theta=0$, and $\theta=-1$. 
  (c) $P_{an}(Q,t)$ vs $Q$; points represent Monte Carlo results (with $L=500$ (red squares) and $L=2000$ (green circles)) while the dashed line is the Poisson distribution  with mean  $\mu=35.4573$ (Eq.~\ref{ac-mut}).
In the tails the agreement improves with increasing $L$.  \label{annealed-ac-ld-causes}} 
\end{figure}

\subsection{Annealed case}

 \noindent The flux of RTPs  depends on the evolution of the particles with an initial position $x(0)$ in $[-v_0 t, 0]$. But what is their number?
 Typically we expect $N \sim \rho v_0 t$, but, in the annealed case,  rare realizations with  large or small $N$ can occur. To capture them
we consider a large interval $[-L,0]$ with $L \gg \rho v_0 t$ and  draw  $L \rho$ initial positions evenly distributed in the interval. Then  only the partciles with  $x(0) > -v_0 t$ are evolved.
   
 \noindent Here we study  RTP with   $\rho=1, v_0=1, \gamma=0.05$. At time $t=400$,  the average flux predicted by Eq.~\ref{ac-mut} is  $\mu=35.4573$,  very close to the one predicted in diffusive limit $\sim \sqrt{t/(2 \gamma \pi)} =35.6825$. Then  typical realizations  ($\theta=0$) are expected to be  similar to the diffusive ones with $D_\text{eff}=v_0^2/(2 \gamma)=10$. When the bias is applied ($\theta =1$) the sampled realizations have a larger flux,  $Q \approx 93 $ and it is  instructive to characterize their statistical properties for RTP and diffusion.

\noindent  In  Fig.~\ref{annealed-ac-ld-causes}(a) we show the initial profile of particles. While for $\theta=0$ it is flat as expected, for $\theta=-1$  it  displays an accumulation of particles around the origin and  total number of particles in the interval $[-v_0 t, 0]$ is much larger than $\rho v_0 t$. In  Fig.~\ref{annealed-ac-ld-causes}(b) we show the histogram (normalized to the number of realizations) of displacement for particles with a positive final position.  The peak for $\theta=0$ has the same location of the peak for $\theta =-1$. The only difference is that more particles have a positive $x(t)$ for $\theta =-1$ than for $\theta=0$.
This suggests that, in the annealed case, larger values of the flux are essentially  due to rare fluctuations of the initial conditions that   are completly insensitive to the nature of the particle motion.

\noindent For this reason the non-gaussian tails of the flux are Poissonian both  for diffusive and active particles.   In Fig.~\ref{annealed-ac-ld-causes}(c) \footnote{Note that to obtain a reliable histogram one has to re-weight the sampled values of $Q$ by a factor $Z(\theta) e^{\theta Q}$.  $Z(\theta)$ is a normalization constant that is determined following the method explained in  \cite{hartmann-epjb-2011, alberto-importance}. } we observe that the agreement between simulations and Poissonian tails increases when $L$ is large, this confirms that the origin of anomalous fluctuations of the flux is in the rare realizations with large initial concentrations of particles close to the origin.  This mechanism cannot work for the quenched case where the initial concentration is always flat.

\subsection{Quenched case}    
In the quenched case the initial condition is fixed and the number of particles with an initial position  in $[-v_0 t, 0]$ is always $N =\rho v_0 t$.
In practice, we fix the position of the first particle
$x_1$ using a uniform random number between $0$ and $-1/(2 \rho)$ and the positions of all the other $N-1$ particles are then slaved to $x_1$ according to
\begin{equation}\label{eqp-inc}
 x_i (0) = x_1 (0) - i,
\end{equation}

\begin{figure}[htb]
\begin{center}
\begin{minipage}{0.4\hsize}
\includegraphics[width=\hsize]{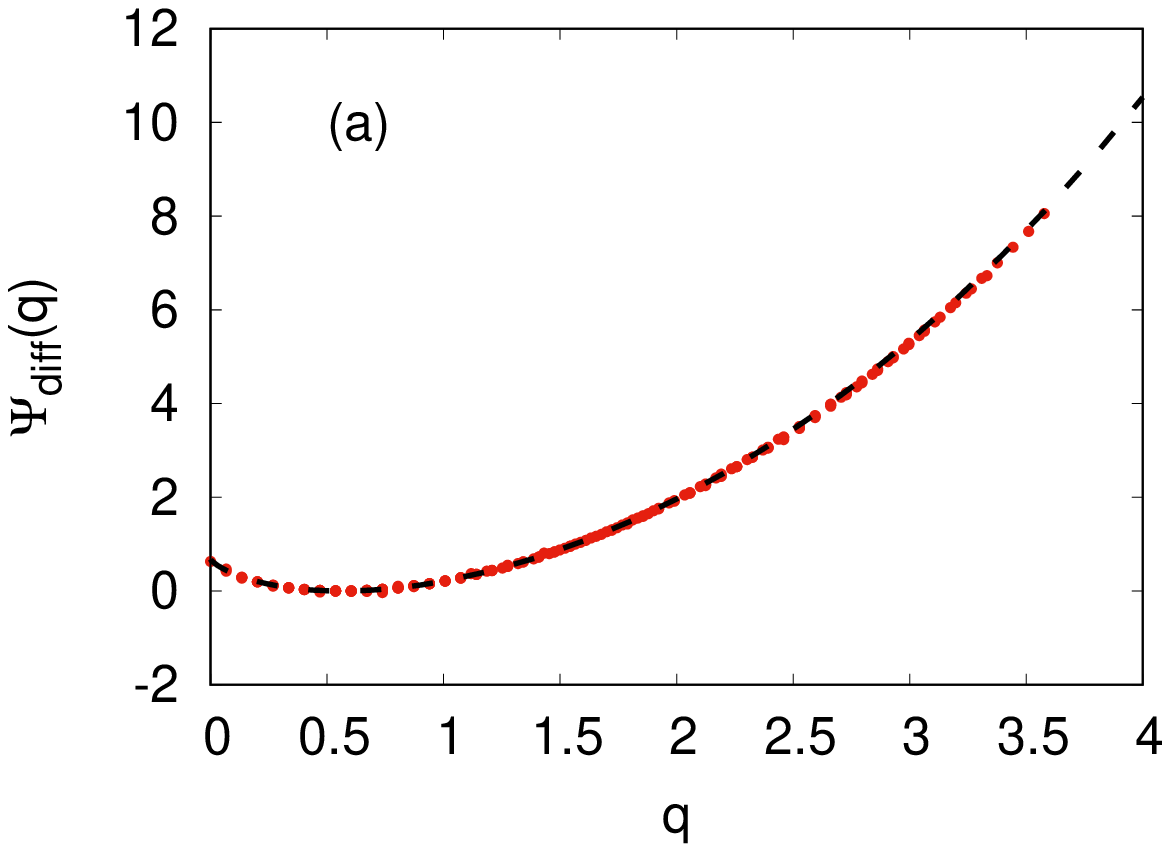}
\end{minipage}
\begin{minipage}{0.4\hsize}
\includegraphics[width=\hsize]{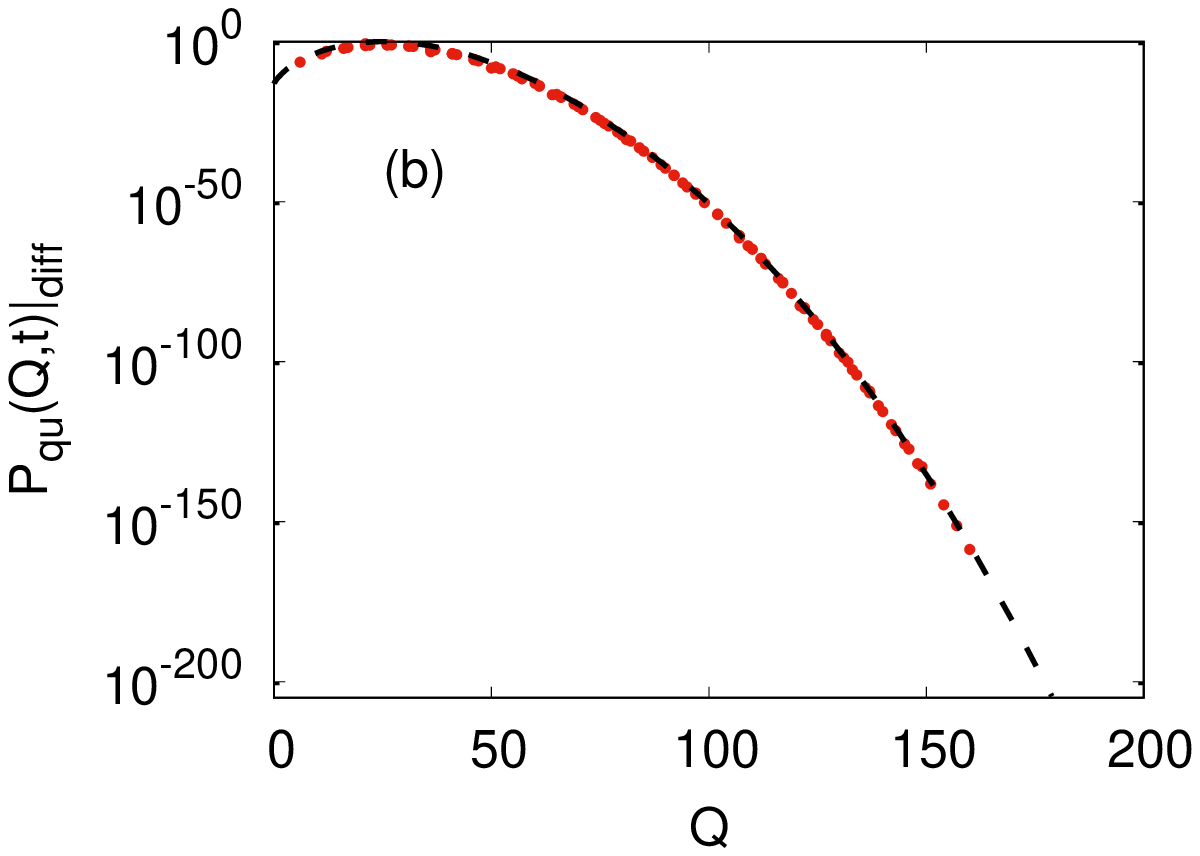}
\end{minipage}
\end{center}
\caption{Quenched case for passive dynamics. Monte Carlo simulations (red points) and exact results obtained from Mathematica (dashed black line) using Eqs.~\ref{phi-1eq} and \ref{fq-bas}.  Simulations were perfomed using  $\rho=1, D=0.2, t=10^4, N=2000$.
(a) $\Psi_{\rm diff}(q)$ vs $q$.
(b) $P_{qu}(Q,t)|_{\rm diff}$ vs $Q$.  The two functions are related via $P_{qu}(Q,t)|_{\rm diff}=\exp[-\rho \sqrt{D t} \Psi_{\rm diff}(\frac{Q}{\rho\sqrt{D t}})] $.
 }
\label{diffusion-quenched-plots}
\end{figure}

\begin{figure}[htb]
\begin{center}
\begin{minipage}{0.4\hsize}
\includegraphics[width=\hsize]{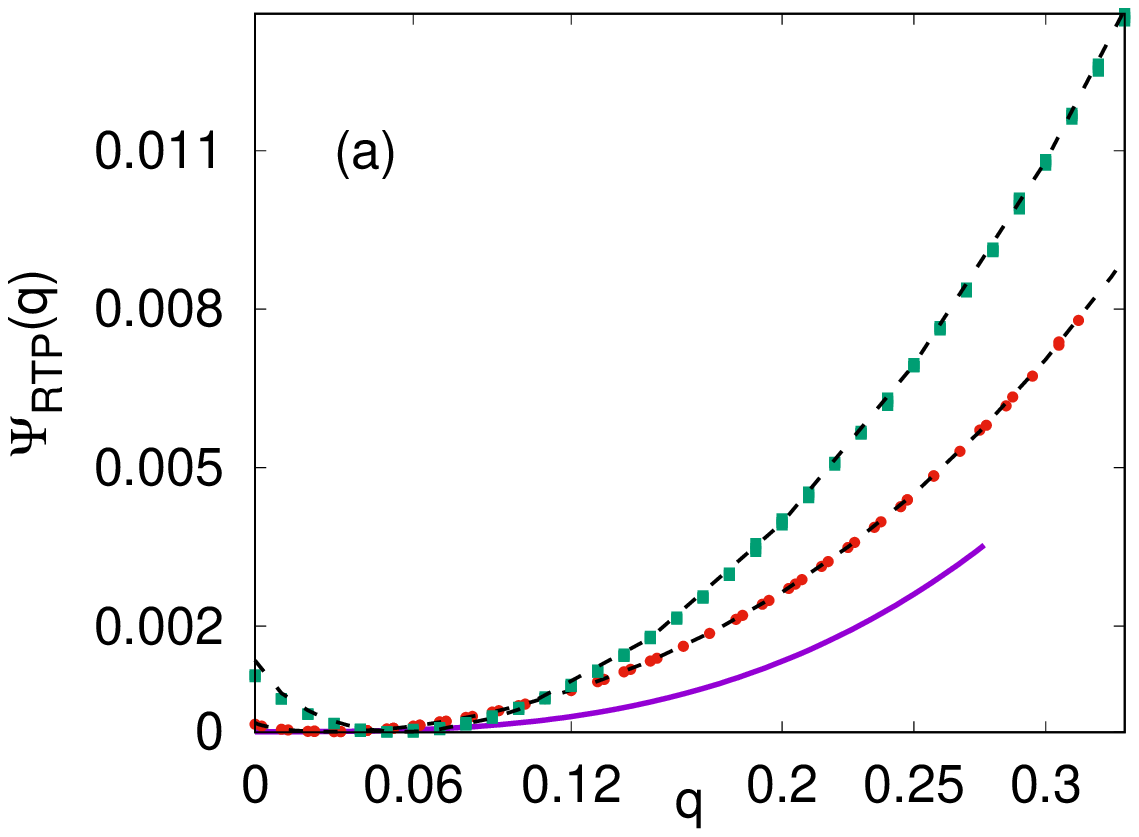}
\end{minipage}
\begin{minipage}{0.4\hsize}
\includegraphics[width=\hsize]{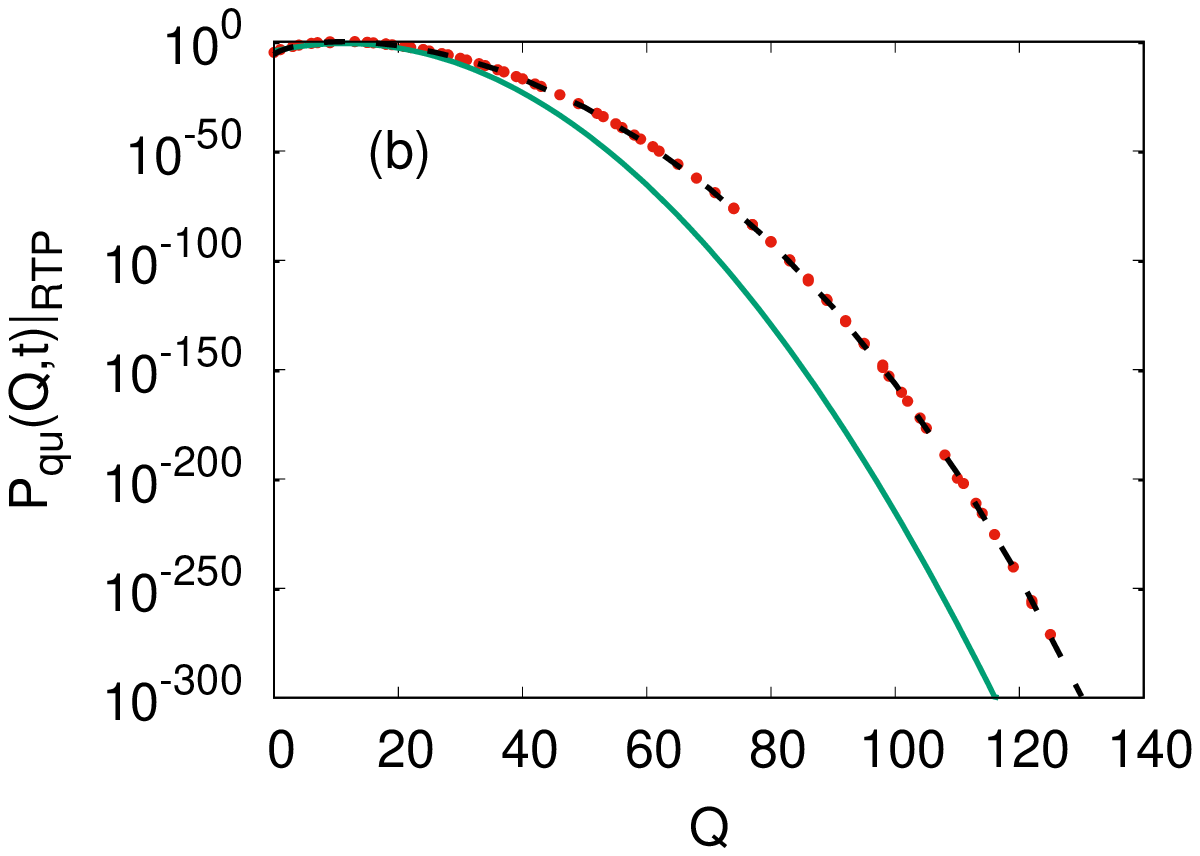}
\end{minipage}
\end{center}
\caption{Quenched case for active dynamics. Monte Carlo simulations (points) and exact results obtained from Mathematica (dashed black line) using Eqs.~\ref{Uzt-exact} and~\ref{def_tildeI}. Simulations were perfomed using  $\rho=v_0=1, \gamma=0.5$ and $t=100$ (green squares) and $t=400$ (red circles). (a)  $\Psi_{\rm RTP}(q)$ vs $q$. The violet solid curve represents $\frac{q^3}{6}$, indicating that $t=400$ is also not large enough to see the ${q^3}$ behavior; also see Fig.~\ref{phi-full-plot}(c) which shows that Eq.~\ref{rtp-ldf-model} becomes a good fit to the exact results only at very large times.
(b)  $P_{qu}(Q,t)|_{\rm RTP}$ vs $Q$ for $t=400$ in the semi-log scale. Note that $P_{qu}(Q,t)|_{\rm RTP}$ decays slower than the Gaussian expected for the typical fluctuations (green solid line).  The two functions are related via
$P_{qu}(Q,t)|_{\rm RTP} = \exp [-\rho v_0 \gamma t^2 \Psi_{\rm RTP}(\frac{Q}{\rho v_0 t})]$.}\label{active-quenched-final-plots}
\end{figure}

We first check the agreement between our Monte Carlo simulations and the analytical predictions.
In Fig.~\ref{diffusion-quenched-plots} we compare
the exact large deviation function $\Psi_{\rm diff}(\frac{Q}{\rho\sqrt{D t}})$ and the exact probability distribution function $P_{qu}(Q,t)|_{\rm diff}$ with the ones obtained from our Monte Carlo simulations. 
For the quenched active case, the results are shown in Fig~\ref{active-quenched-final-plots} where we also plot the large deviation function $\Psi_{\rm RTP}(\frac{Q}{\rho v_0 t})$, using for the Monte Carlo data the definition $\Psi_{\rm RTP}(q) = -\frac{{\rm log} P_{\rm qu}(Q,t)}{\rho v_0 \gamma t^2}$ with $q=\frac{Q}{\rho v_0 t}$. We note that the Monte Carlo results perfectly match the predictions for both RTP and diffusion.

\begin{figure}[htb]
\begin{center}
\begin{minipage}{0.4\hsize}
\includegraphics[width=\hsize]{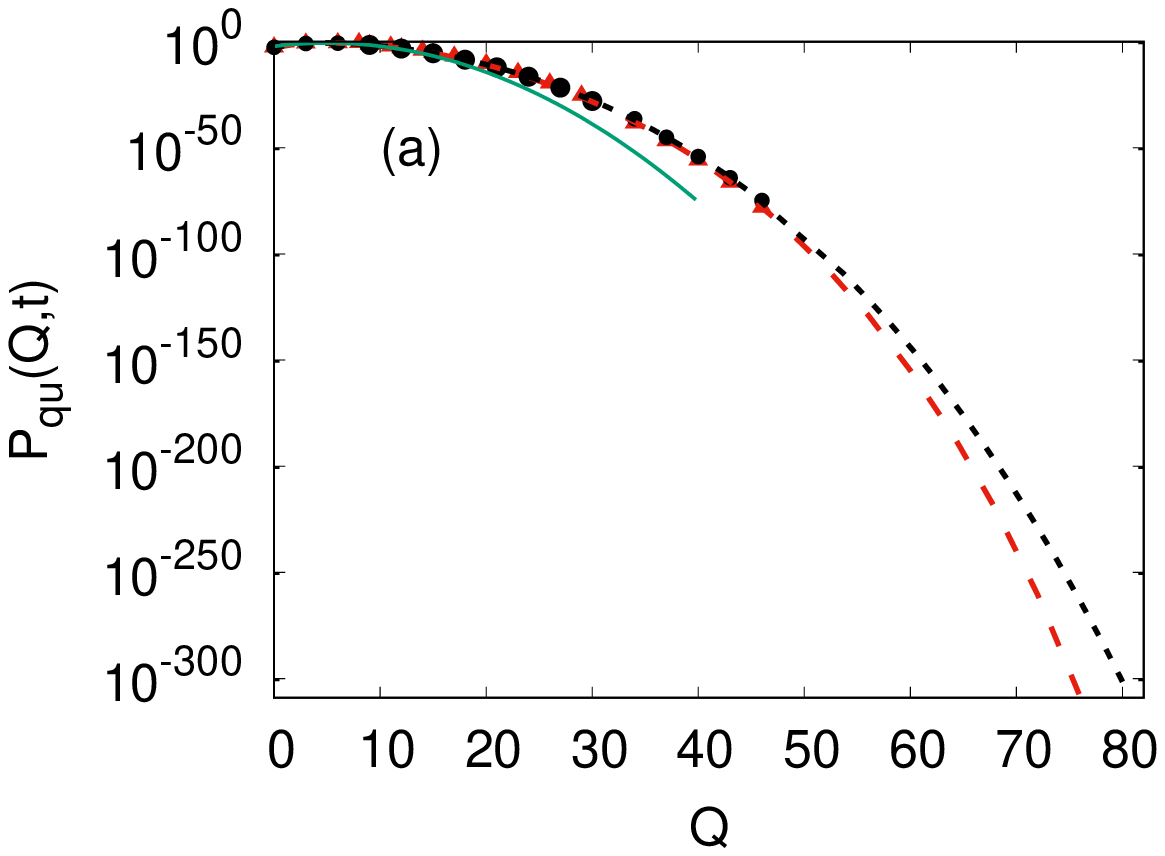}
\end{minipage}
\begin{minipage}{0.4\hsize}
\includegraphics[width=\hsize]{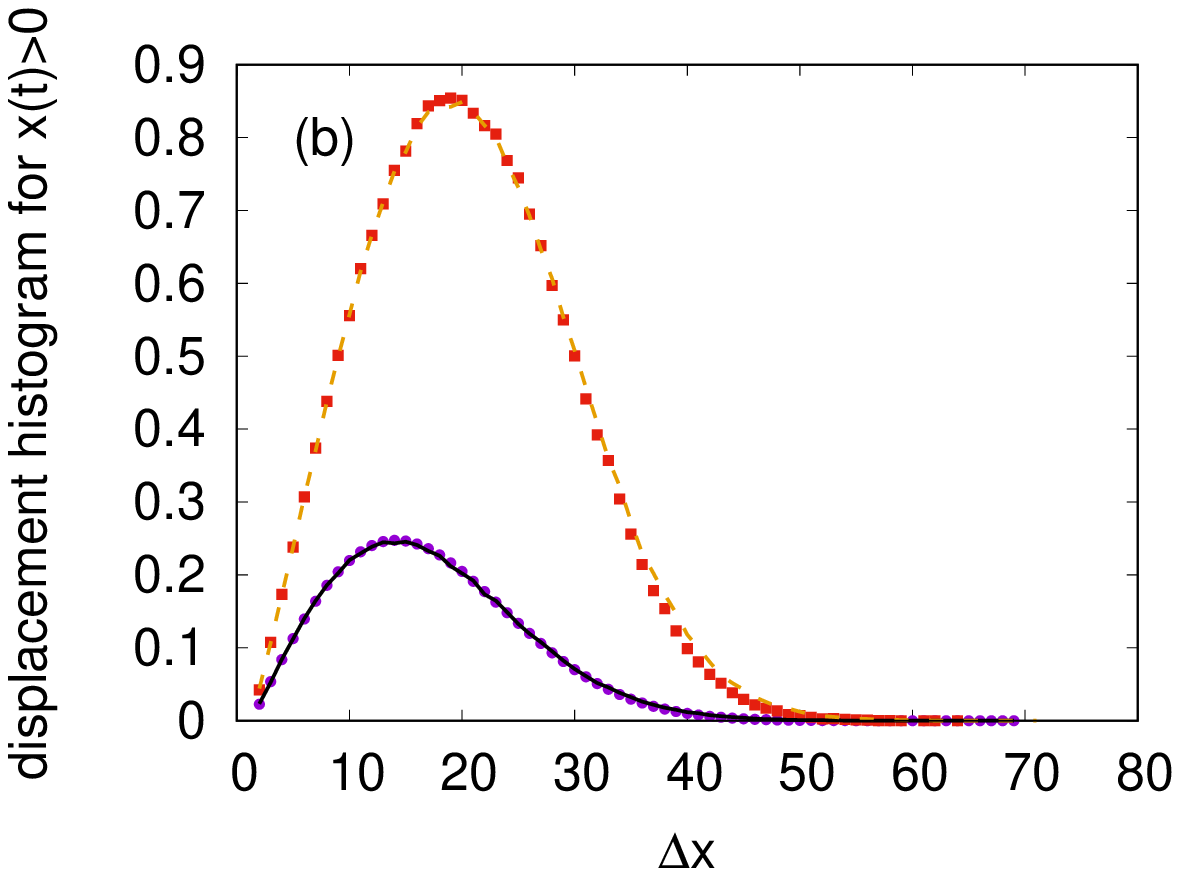}
\end{minipage}
\end{center}
\caption{ (a) $P_{\rm qu}(Q,t)$ vs $Q$. RTPs  with  $\rho=v_0=1, \gamma=0.5$ and $t=80$ (long-dashed red lines, triangles) are compared with  the diffusive ones with $D_\text{eff}=v_0^2/(2 \gamma)=1$ (short-dashed black lines, circles). Exact results (lines) and Monte Carlo simulations (points). The solid green line represents the Gaussian valid at small $Q$.
(b) Histogram  (normalized to the number of realizations) of the displacement for particles with a positive final position with $\theta=0$ (violet circles for RTP/ black -solid lines for diffusion), and $\theta=-3$ (red squares for RTP/ yellow-dashed lines for diffusion).}
\label{comp-rtp-diff-gmt40}
\end{figure}

The difference between RTP and diffusive behaviour in the quenched case is shown is Fig.~\ref{comp-rtp-diff-gmt40} (a). There we compare the exact distribution of RTP with $\rho=v_0=1, \gamma=0.5$ and $t=80$ (red)  with  the diffusive one with $D_\text{eff}=v_0^2/(2 \gamma)=1$.
 Both distributions  deviate from the Gaussian (solid green) but they become separable from each other only at extremely small probabilities, when $Q \simeq 60 \approx \rho v_0 t$. The Importance Sampling strategy  enables us to explore the non-Guassian tails of the distribution but not the extremely rare configurations where the finerprints of the activity are present. Indeed in  Fig.~\ref{comp-rtp-diff-gmt40} (b) we can hardly see a difference in histogram of the displacement of particles with positive final position between diffusion and RTP, both for $\theta =0$ and $\theta=-3$. This is because the displacements involved are still very small compare to $v_0t =80$.

\section{Conclusion}\label{conclu}

In this paper, we have presented a general framework to study current fluctuations for non-interacting particles executing a common random dynamics in one dimension and starting from a step initial condition. The probability distribution $P(Q,t, \{x_i\})$ depends on the initial positions of the particles $\{x_i\}<0$. The initial positions are distributed uniformly on the negative axis with a uniform density $\rho$. There are two different ways to perform the average over the initial positions, namely (i) annealed and (ii) quenched averages, in analogy with disordered systems: here the initial condition plays the role of the disorder. In the annealed case, the distribution $P(Q,t,\{x_i\})$ is averaged directly over the initial positions. In contrast, in the quenched case, one considers the configurations of $x_i$'s that lead to the most likely current distribution (i.e., the {\it typical} current distribution). In both cases, we have shown that, for noninteracting particles, the distribution can be fully characterized in terms of the single particle Green's function, which in general will depend on the dynamics of the particles. In this article, we have focused mostly on two different dynamics: a) when the single particle undergoes simple diffusion and b) when the single particle undergoes run-and-tumble dynamics (RTP). 

For the annealed case, we have shown that $P_{\rm an}(Q,t)$, at all times, is given by a Poisson distribution, with parameter $\mu(t)$ given by the exact formula in Eq.~(\ref{muan-mod-def}). We provide exact formula for $\mu(t)$ in the RTP case (for the diffusive case this was known already
from Ref. \cite{derrida-gers}). For the quenched diffusive case we show that our formalism correctly recovers the large deviation result obtained in Ref. \cite{derrida-gers}) using a different approach. For the RTP case, we showed that there is a new large deviation regime with $Q \sim t$, where $P_{\rm qu}(Q,t) \sim \exp\left[-\rho \, v_0 \, \gamma \, t^2 \, \Psi_{\rm RTP}\left(\frac{Q}{\rho \, v_0 \,t}\right)\right]$. One of the main results of this paper is an explicit computation of the rate function $\Psi_{\rm RTP}(q)$ given by
\begin{equation}\label{rtp-ldf-model_conclusion}
\Psi_{\rm RTP}(q)=q-\frac{q}{2}\sqrt{1-q^2}-{\rm sin}^{-1}\left[ \sqrt{\frac{1-\sqrt{1-q^2}}{2}} \right]\;, \quad 0 \leq q \leq 1 \;.
\end{equation}

Our method gives access to another physical observable, namely the probability of an extremely rare event that 
there is no particle on the right side of the origin at time $t$. We have shown that this is just the probability of having
zero flux up to time $t$, i.e. $P(Q=0,t)$, both for the annealed and the quenched case. For the annealed case, this is just
$P_{\rm an}(Q=0,t) = e^{-\mu(t)}$. For the quenched case, we have that, both for the diffusive and RTP cases, this probability 
decays at late times as a stretched exponential $P_{\rm qu}(Q=0,t) \sim e^{-\bar{\alpha} \sqrt{D_{\rm eff}\,t}}$, where we computed the constant $\bar{\alpha} = 0.675336\ldots$ analytically [see Eq. (\ref{C})]. For diffusive particles, $D_{\rm eff} = D$ while for RTP's, $D_{\rm eff} = v_0^2/(2 \gamma)$. 

We have also verified our analytical predictions by numerical simulations. Computing numerically the large deviation function is far from trivial. Even for the diffusive case the large deviation function predicted for $P(Q,t)$ (both annealed and quenched) in Ref. \cite{derrida-gers} was
never verified numerically. In this paper, we used a sophisticated importance sampling method to compute numerically this large deviation 
function in the diffusive case up to an impressive accuracy of order $10^{-200}$. We further used the same technique to compute the large deviation function in the RTP case.  

The formalism developed in this paper can be easily generalized in different directions. For instance, one can compute the flux distribution exactly for the case where there are, initially, arbitrary densities $\rho_{\rm left}$ and $\rho_{\rm right}$ to the left and to the right of the origin respectively, both the diffusive and for the RTP cases. One could also generalise this result in higher dimensions, with step-like initial conditions, where for instance one region of the space is initially occupied by particles with uniform density. For the diffusive case, the flux distribution in the presence of hard-core repulsions between particles was studied in Ref. \cite{derrida-gers-sep} (for the simple symmetric exclusion process). It would be interesting to see whether our formalism can be generalized to study the flux distribution for RTP's with hard core repulsions.

\begin{acknowledgments}
We acknowledge support from the project 5604-2 of the Indo-French Centre for the Promotion of Advanced Research (IFCPAR). 
\end{acknowledgments}

\end{document}